\documentclass[12pt]{iopart}
\usepackage{amsfonts}
\usepackage{amssymb}
\usepackage{graphicx}
\usepackage{setspace}
\usepackage{CJK}
\usepackage{moresize}
\usepackage[utf8]{inputenc}
\usepackage[english]{babel}

\usepackage{booktabs}
\usepackage{siunitx}
\usepackage{booktabs, makecell, multirow}

\newcommand{\begM}{\begin{multline}}
\newcommand{\eM}{\end{multline}}

\def\b{\beta}
\def\a{\alpha}
\newcommand{\m}{\mu}
\newcommand{\n}{\nu}
\newcommand{\p}{\partial}

\def\){\Big)}
\def\({\Big(}

\def\r{\rho}

\def\k{\kappa}
\def\t{\tau}
\def\g{\gamma}
\def\x{\chi}

\begin{document}

\title[IJMPD Manuscript Part II: Complete dual formulation]{A connection between linearized Gauss-Bonnet gravity and classical electrodynamics II: Complete dual formulation}
\date{\today}

\linespread{1.25}

\author{Mark Robert Baker$^{1,2}$}

\address{$^1$ Department of Physics and Astronomy, University of Western Ontario, London, ON, N6A 3K7, Canada}

\address{$^2$ The Rotman Institute of Philosophy, University of Western Ontario, London, ON, N6A 5B7, Canada}

\ead{mbaker66@uwo.ca}

\vspace{10pt}

\begin{indented}
\item[]02 Apr 2021
\end{indented}

\begin{abstract}
In a recent publication a procedure was developed which can be used to derive completely gauge invariant models from general Lagrangian densities with $N$ order of derivatives and $M$ rank of tensor potential. This procedure was then used to show that unique models follow for each order, namely classical electrodynamics for $N = M = 1$ and linearized Gauss-Bonnet gravity for $N = M = 2$. In this article, the nature of the connection between these two well explored physical models is further investigated by means of an additional common property; a complete dual formulation. First we give a review of Gauss-Bonnet gravity and the dual formulation of classical electrodynamics. The dual formulation of linearized Gauss-Bonnet gravity is then developed. It is shown that the dual formulation of linearized Gauss-Bonnet gravity is analogous to the homogenous half of Maxwell's theory; both have equations of motion corresponding to the (second) Bianchi identity, built from the dual form of their respective field strength tensors. In order to have a dually symmetric counterpart analogous to the non-homogenous half of Maxwell's theory, the first invariant derived from the procedure in $N = M = 2$ can be introduced. The complete gauge invariance of a model with respect to Noether's first theorem, and not just the equation of motion, is a necessary condition for this dual formulation. We show that this result can be generalized to the higher spin gauge theories, where the spin-$n$ curvature tensors for all $N = M = n$ are the field strength tensors for each $n$. These completely gauge invariant models correspond to the Maxwell-like higher spin gauge theories whose equations of motion have been well explored in the literature.
\end{abstract}

\maketitle

\section{Motivation}

\normalsize

In \cite{baker2019}, a procedure for deriving completely gauge invariant models from general linear combinations of derivatives of order $N$ and rank of potential $M$ was developed. Complete gauge invariance occurs for a model when the Lagrangian density, equation of motion and energy-momentum tensor are all independently and exactly gauge invariant. The procedure involves solving for the free coefficients in the linear combination with respect to Noether's (first) theorem \cite{noether1918,kosmann2011noether} such that the model is completely gauge invariant under a particular gauge transformation. In the case of $N = M = 1$ under a spin-1 gauge transformation, electrodynamics is uniquely derived from the procedure. In the case of $N = M = 2$, under a spin-2 gauge transformation (sometimes referred to as linearized diffeomorphisms), linearized Gauss-Bonnet gravity is uniquely derived. This result has since been used to prove that the Noether and Hilbert energy–momentum tensor are not, in general, equivalent \cite{baker2021a}.

The connection of these models to a common procedure raised an obvious question, what is the reason for this connection, and why Gauss-Bonnet gravity of the many metric theories of gravity that exist in the literature. The present article attempts to answer both questions through the Gauss-Bonnet theorem, and the additional non-trivial property shared by these two models, complete dual formulation of the Lagrangian, equation of motion and energy-momentum tensor. Once again these models will be derived from the Noether identity from Noether's first theorem, given below for a general potential $\Phi_A$ \cite{baker2019},

\begin{equation}
\fl
\eqalign{ \left( \frac{\partial \mathcal{L}}{\partial \Phi_A}
 - \partial_\mu \frac{\partial \mathcal{L}}{\partial (\partial_\mu \Phi_A)} 
 + \partial_\mu \partial_\omega \frac{\partial \mathcal{L}}{\partial (\partial_\mu \partial_\omega \Phi_A)} + \dots \ \right) \delta \Phi_A
 \\
+ \partial_\mu \left(  \eta^{\mu\nu} \mathcal{L} \delta x_\nu
+ \frac{\partial \mathcal{L}}{\partial (\partial_\mu \Phi_A)} \delta \Phi_A 
+ \frac{\partial \mathcal{L}}{\partial (\partial_\mu \partial_\omega \Phi_A)} \partial_\omega \delta \Phi_A
- \left[ \partial_\omega \frac{\partial \mathcal{L}}{\partial (\partial_\mu \partial_\omega \Phi_A)} \right] \delta \Phi_A
+ ... \ \right) = 0 \ . }
   \label{genenergy}
\end{equation}

The article will be structured as follows. In Section \ref{s2} an overview of Gauss-Bonnet gravity is given, with its connection to the Gauss-Bonnet theorem, and how it is derived from the Euler class $e(\Omega)$ in the integrand of the Gauss-Bonnet theorem. Next the dual linearized Riemann tensor is introduced and connected to the original results of Lanczos that first noted these connections between the dual Riemann tensors, their scalars, and what is now known as the Gauss-Bonnet Lagrangian. In Section \ref{s3} an overview of the dual formulation of electrodynamics is presented, and how every component of the theory with respect to Noether's theorem (the Lagrangian density, equation of motion and energy-momentum tensor) can be expressed explicitly in dual form.

Section \ref{s4} is dedicated to converting the linearized Gauss-Bonnet gravity model from \cite{baker2019} into dual form with respect to Noether's first theorem. It is shown that this model has the same general dual formulation as the homogenous half of Maxwell's theory; the equation of motion is the second Bianchi identity built from the dual linearized Riemann tensor tensor. In Section \ref{s5} possible invariants derived from the procedure in \cite{baker2019} are discussed that can give complete dual formulation analogous to the complete dual formulation of electrodynamics. Indeed the invariant $R_{\mu\nu\alpha\beta} R^{\mu\nu\alpha\beta}$ yields this with respect to the Lagrangian and equation of motion, but complications arise with the third term in the energy-momentum tensor from Noether's first theorem. This is because, as shown in \cite{baker2019}, only for very particular Lagrangian densities can this be made gauge invariant and symmetric, namely the Gauss-Bonnet combination. Two possible remedies to this problem are given and possible ramifications are discussed. 

In Section \ref{s6} the internal dual formulations of the respective electodynamics and linearized gravity models are presented. Section \ref{s7} gives the general forms for the analogous expressions between the two models which are generalized for any independently gauge invariant spin-$n$ field strength tensor. These models are the Maxwell-like higher spin gauge theories for the spin-$n$ curvature tensors \cite{francia2002,francia2012,bekaert2015}. The complete dual and gauge invariance of these models with respect to Noether's theorem provides more compelling evidence for the requirement of complete invariance properties of physical theories with respect to all components: the Lagrangian density, equation of motion and energy-momentum tensor of the model.

\section{Gauss-Bonnet gravity and the dual Riemann tensors \label{s2}}

We begin by providing details regarding the origin of Gauss-Bonnet gravity, its relation to the Gauss-Bonnet theorem, and how the common Lagrangian in the literature is obtained from the Euler class in the integrand of the theorem. This is necessary because the required calculations for our article are scattered throughout the literature, if at all. The book by Eguchi, Gilkey and Hanson \cite{eguchi1980} will be taken as the primary reference for details here, however even this reference is missing considerable detail and explanation. The Gauss-Bonnet theorem got its name by the work of Gauss in 1827 (Gauss's theorem egregrium) \cite{gauss1827} and Bonnet in 1848 \cite{bonnet1848}, although neither of these presentations are what we refer to as the Gauss-Bonnet theorem in the present day (they were earlier developments of the theorem). The modern day version was first presented by Dyck in 1890 \cite{dyck1890} for the specific case of $R^3$, and finally to $n$ dimensions by Hopf in 1926 \cite{hopf1926}, with the proof of the general formula for Riemannian manifolds being completed by Chern \cite{chern1944}. 

It was Allendoerfer \cite{allendoerfer1940} who first showed that the integrand for a Riemannian manifold of dimension $d$ is the general expression of which the special case $d=4$ is what we will derive below (the Gauss-Bonnet Lagrangian). This was recognized for $d=4$ indirectly by Lanczos \cite{lanczos1938} a couple years earlier, but from motivations discussed later in this section. A more detailed account of this history was given by \cite{wu2008}.  The modern form of the Gauss-Bonnet theorem is sometimes referred to as the generalized Gauss-Bonnet theorem or Chern-Gauss-Bonnet theorem to make a distinction between the more advanced (modern) version compared to the work of Gauss and Bonnet. The modern form of the theorem states,

\begin{equation}
\chi(M) = \int_{\bar{M}} e(\Omega),  \label{cherngaussbonnettheorem}
\end{equation}

where $\chi$ is the Euler characteristic of manifold $\bar{M}$ and $e(\Omega)$ is the Euler class.  The Euler class can be expressed in terms of the Pfaffian of the curvature form $Pf(\Omega)$,

\begin{equation}
e(\Omega) = \frac{1}{(2\pi)^{d/2}} Pf(\Omega).  \label{eulerclass}
\end{equation}

The Pfaffian of the curvature form for a Riemannian manifold in 4 dimensions (4D) is the $d=4$ case. The expression for the Euler class of the curvature form in this case was first given by Allendoerfer \cite{allendoerfer1940}, but in more explicit notation by Eguchi, Gilkey and Hanson \cite{eguchi1980}. The Pfaffian is given by,

\begin{equation}
Pf(\Omega) = \frac{1}{8} \epsilon_{abcd} \Omega^{ab} \wedge \Omega^{cd},
\end{equation}

where the curvature 2-form for the Riemannian manifold is given in terms of the Riemann tensor, $\Omega^{\mu\nu} = \frac{1}{2} R^{ \mu\nu}_{\ \ \rho\sigma} dx^\rho \wedge dx^\sigma $. The Euler class therefore reads,

\begin{equation}
e(\Omega) = \frac{1}{4} \frac{1}{32 \pi^2} \epsilon_{\mu\nu\alpha\beta} R^{ \mu\nu}_{\ \ \rho\sigma} R^{ \alpha\beta}_{\ \ \lambda\gamma} dx^\rho \wedge dx^\sigma \wedge  dx^\lambda \wedge dx^\gamma .
\end{equation}

Expanding out this summation yields,

\begin{equation}
\fl
\eqalign{
e(\Omega) =  \frac{1}{32 \pi^2}  [8 
(R^{ 12}_{\ \ 34} R^{ 34}_{\ \ 12} 
+   R^{ 14}_{\ \ 23} R^{ 23}_{\ \ 14} 
+ R^{ 13}_{\ \ 24} R^{ 24}_{\ \ 13}  
-  R^{ 12}_{\ \ 13} R^{ 34}_{\ \ 24} 
+ R^{ 12}_{\ \ 14} R^{ 34}_{\ \ 23} 
+ R^{ 12}_{\ \ 23} R^{ 34}_{\ \ 14} 
\\
- R^{ 12}_{\ \ 24} R^{ 34}_{\ \ 13} 
+ R^{ 13}_{\ \ 12} R^{ 42}_{\ \ 34}  
+ R^{ 13}_{\ \ 14} R^{ 42}_{\ \ 23}  
+ R^{ 13}_{\ \ 23} R^{ 42}_{\ \ 14}  
+ R^{ 13}_{\ \ 34} R^{ 42}_{\ \ 12}  
+   R^{ 14}_{\ \ 12} R^{ 23}_{\ \ 34} 
-   R^{ 14}_{\ \ 13} R^{ 23}_{\ \ 24} 
\\
-   R^{ 14}_{\ \ 24} R^{ 23}_{\ \ 13} 
+   R^{ 14}_{\ \ 34} R^{ 23}_{\ \ 12} 
+ R^{ 12}_{\ \ 12} R^{ 34}_{\ \ 34} 
+ R^{ 13}_{\ \ 13} R^{ 24}_{\ \ 24}  
+   R^{ 14}_{\ \ 14} R^{ 23}_{\ \ 23} )
   ] 
dx^1 \wedge dx^2 \wedge  dx^3 \wedge dx^4 .
} 
\end{equation}

What is in square brackets above is identically $[8(\dots)] = R_{\mu\nu\alpha\beta} R^{\mu\nu\alpha\beta} - 4 R_{\mu\nu} R^{\mu\nu} + R^2$. Additionally, using the relationship $d^4 x = dx^1 \wedge dx^2 \wedge dx^3 \wedge dx^4 = \frac{1}{24} \epsilon_{\mu\nu\alpha\beta} dx^\mu \wedge dx^\nu \wedge dx^\alpha \wedge dx^\beta$, the Euler class for the Riemannian manifold in 4D is,

\begin{equation}
e(\Omega) = \frac{1}{32 \pi^2} (R_{\mu\nu\alpha\beta} R^{\mu\nu\alpha\beta} - 4 R_{\mu\nu} R^{\mu\nu} + R^2) d^4 x .
\end{equation}

Therefore the Gauss-Bonnet theorem for this case reads $\chi(M) = \frac{1}{32 \pi^2} \int_{\bar{M}} (R_{\mu\nu\alpha\beta} R^{\mu\nu\alpha\beta} - 4 R_{\mu\nu} R^{\mu\nu} + R^2) d^4 x$. It is this integrand of the Gauss-Bonnet theorem in 4D that is precisely the Lagrangian density for the Gauss-Bonnet gravity model. 

This contribution was first introduced to the physics community by Cornelius Lanczos in 1938 \cite{lanczos1938}, although happened across by very different means. Lanczos was considering various invariants that can be obtained from the Riemannian tensors, as presented in his paper $I_1 = R_{\mu\nu} R^{\mu\nu}$, $I_2 = R^2$ and $I_3 = R_{\mu\nu\alpha\beta} R^{\mu\nu\alpha\beta}$. Of course these are the 3 invariants found above in the Gauss-Bonnet theorem, and the 3 invariants derived in \cite{baker2019} as $\mathcal{L} = \tilde{a} R_{\mu\nu\alpha\beta} R^{\mu\nu\alpha\beta} + \tilde{b} R_{\mu\nu} R^{\mu\nu} + \tilde{c} R^2$, where $\tilde{a} = \frac{1}{4}$, $\tilde{b} = -1$ and $\tilde{c} = \frac{1}{4}$ for linearized Gauss-Bonnet gravity (Equation (\ref{LGBLAG1})).

 What Lanczos noticed is that if we consider a Lagrangian density formed from the combination $I_3 - 4 I_1 + I_2$, it will make no contribution to the equation of motion. This result is now more appropriately understood as the nature of topological invariants, which can be expressed as a total derivative in the action. This result, however, as emphasized in \cite{baker2019}, does not mean that the action will not contribute to the energy-momentum tensor of the model. In addition, as we show in Section \ref{s4}, the equation of motion (while zero) is in fact the second Bianchi identity analogous to the homogenous half of Maxwell's equations. The precise form of the energy-momentum tensor is a well known expression to string theorists for several decades \cite{ray1978,zwiebach1985,boulware1985,myers1987}, as derived for linearized Gauss-Bonnet gravity from Noether's first theorem in \cite{baker2019},

\begin{equation}
\fl
T^{\omega\nu} = -  R^{\omega\rho\lambda\sigma} R^\nu_{\ \rho\lambda\sigma}
 + 2  R_{\rho\sigma} R^{\omega \rho \nu \sigma}
+ 2  R^{\omega\lambda} R^\nu_{\ \lambda}
 -  R R^{\omega\nu} + \frac{1}{4} \eta^{\omega\nu}(R_{\mu\lambda\alpha\beta} R^{\mu\lambda\alpha\beta} - 4 R_{\mu\gamma} R^{\mu\gamma} + R^2) . \label{gaussemt}
\end{equation}

Gauss-Bonnet gravity is an extensively published model in the literature, interests which have only been increasing in recent years  \cite{charmousis2002,cherubini2002,granda2012,marugame2016,benetti2018,glavan2020}. This past year \cite{glavan2020} has attracted significant attention in the literature by claiming the the Gauss-Bonnet model can be used to predict `new' gravitational dynamics solving which can explain several still unexplained phenomena. We note that many authors have been writing to support, criticize and further this result \cite{fernandes2020}.

Lanczos did consider two additional invariants built from dual tensors \cite{lanczos1938}, 

\begin{equation}
{\bf{R}}_{\alpha\beta\mu\nu} = \frac{1}{2} R^{\rho\sigma}_{\ \ \mu\nu} \epsilon_{\rho\sigma\alpha\beta} , \label{Riesingdual}
\end{equation}

\begin{equation}
\mathcal{R}_{\mu\nu\alpha\beta} = \frac{1}{4} R^{\rho\sigma\lambda\gamma} \epsilon_{\rho\sigma\mu\nu} \epsilon_{\alpha\beta\lambda\gamma} , \label{Riedoubdual}
\end{equation}

which he called `simply' dual ${\bf{R}}^{\alpha\beta\mu\nu}$ and `doubly' dual $\mathcal{R}^{\alpha\beta\mu\nu}$, respectively. The two invariants he considered were each of these contracted with the Riemann tensor, $K_1 = {\bf{R}}^{\alpha\beta\mu\nu} R_{\alpha\beta\mu\nu}$ and $K_2 = \mathcal{R}^{\alpha\beta\mu\nu} R_{\alpha\beta\mu\nu}$. From this he showed that the invariant $K_2$ can be used to express the combination which makes no contribution to the equation of motion $K_2 = I_3 - 4 I_1 + I_2$. This will be the starting point for Section \ref{s4} where the linearized Gauss-Bonnet model will be completely rewritten into an explicit dual formulation, as in the case of dual electrodynamics. In order to do this, Section \ref{s3} will first present the complete electrodynamics model in dual form.

\section{Dual electrodynamics \label{s3}}

\subsection{Dual electrodynamic scalars}

The dual formulation of electrodynamics has an interesting history. Heaviside first noticed the dual invariance of the complete 8 Maxwell equations when he first wrote them in vector form \cite{heaviside1894}. Maxwell's equations were presented from the both the field strength tensor $F^{\mu\nu}$ (the 4 non-homogenous equations in Equation (\ref{MaxEOMMNH})) and dual tensor $\mathcal{F}^{\mu\nu}$ (the 4 homogenous equations in Equation (\ref{mheomdual})) by Minkowski in \cite{minkowski1909}. Later the complete 8 Maxwell equations were reformulated into a single field strength tensor $F^{\mu\nu}$ by Einstein \cite{einstein1916}. The appeal here was that a single field strength tensor $F^{\mu\nu}$ could be defined from which all of Maxwell's equations could be presented. The downside was that explicit dual invariance of the model was hidden as a consequence, and that the homogenous half of Maxwell's equations were presented simply as a property of the field strength tensor (from the second Bianchi identity $\partial_\sigma F_{\alpha\beta}+\partial_\beta F_{\sigma\alpha}+\partial_\alpha F_{\beta\sigma} = 0$), rather than following in the Euler-Lagrange equation from a fundamental Lagrangian density. 

Since this time, the Lagrangian density considered to be fundamental to electrodynamics is $\mathcal{L} = - \frac{1}{4} F_{\alpha\beta} F^{\alpha\beta}$ as derived from the procedure in [1], where $F_{\alpha\beta} = \partial_\alpha A_\beta - \partial_\beta A_\alpha$. This Lagrangian density yields the non-homogenous half of Maxwell's equations in Equation (\ref{MaxEOMMNH}) from the Euler-Lagrange equation. Considering the dual tensor of electrodynamics $\mathcal{F}_{\mu\nu} =  \frac{1}{2} \epsilon_{\mu\nu\rho\sigma} {F}^{\rho\sigma}  $, it is possible build in principle 3 invariants, $M_1 = F_{\alpha\beta} F^{\alpha\beta}$, $M_2 = \mathcal{F}_{\mu\nu} \mathcal{F}^{\mu\nu}$ and $M_3 = {F}_{\mu\nu} \mathcal{F}^{\mu\nu}$. It is well known that the first two can be expressed in terms of one another as $M_1 = - M_2$. The common objection to $M_3$ is that it should not be included in the action because it can change sign under an odd numbered parity transformation, since it is formed from the inner product of polar vector $\vec{E}$ and axial vector $\vec{B}$. This sign change however, does not effect the equation of motion. Since Lagrangians which are not exactly gauge invariant but admit gauge invariant equations of motion (such as the spin-2 Fierz-Pauli action invariant up to a surface term \cite{baker2019,padmanabhan2008,magnano2002,dewit1980}) are well accepted in the literature, a change of sign is also a negligible problem if it does not affect the physical model. We will not focus on this philosophical question in this article. 

Note that the authors of a highly cited paper on dual electrodynamics \cite{cameron2012,bliokh2013} propose a new Lagrangian of the form $M_1 + M_2$ where the dual field strength is redefined as $\mathcal{F}_{\mu\nu} = \partial_\mu C_\nu - \partial_\nu C_\mu$ in terms of a second 4-potential $C_\mu$. This has been proposed by numerous other authors throughout the years without gaining much traction. From this perspective the basic idea is that variation with respect to both $A_\mu$ and $C_\mu$ of the Lagrangian $M_1 + M_2$ will yield all 8 of Maxwell's equations, the 4 non-homogenous equations from variation with respect to $A_\mu$ and the 4 homogenous equations from variation with respect to $C_\mu$. This differs from the procedure in \cite{baker2019} and that general view that electrodynamics is built from a single potential $A_\mu$, so the presentation in \cite{cameron2012,bliokh2013} will not be considered here. If the general Lagrangian density in \cite{baker2019} is built using both potentials in separate terms, with the gauge transformation $C_\mu' = C_\mu + \partial_\mu \phi$, then their presentation can also be derived. Their thesis, however, that electrodynamics should be conventionally expressed in dual invariant form to (i) obtain all of Maxwell's equation from the variational approach (since $\mathcal{L} = - \frac{1}{4} F_{\alpha\beta} F^{\alpha\beta}$ only yields the non-homogenous equations), and (ii) allow for a more symmetric presentation of all conservation laws, is hard to argue. The following presentation of the dual formulation is perhaps superior given the complete derivation of Maxwell's theory from $M_1,M_2,M_3$ without the need to introduce any non-canonical potential vectors \cite{gibbons1995,gaillard1998,kuzenko2013}.

\subsection{Generalized Kronecker delta}

In order to perform many of the calculations involving 4D dual expressions for classical electrodynamics and the linearized gravity models discussed in this article, it is necessary to review the generalized Kronecker delta in 4D for a Minkowski spacetime. The generalized Kronecker delta is defined as the determinant of the Kronecker deltas of the permuted indices as follows,

\begin{equation}
\delta_{\rho\sigma\mu\nu}^{\alpha\beta\lambda\gamma} = 
\left|
\pmatrix{
  \delta^\alpha_\rho & \delta^\beta_\rho & \delta^\lambda_\rho & \delta^\gamma_\rho \cr
  \delta^\alpha_\sigma & \delta^\beta_\sigma & \delta^\lambda_\sigma & \delta^\gamma_\sigma \cr
  \delta^\alpha_\mu & \delta^\beta_\mu & \delta^\lambda_\mu & \delta^\gamma_\mu \cr
  \delta^\alpha_\nu & \delta^\beta_\nu & \delta^\lambda_\nu & \delta^\gamma_\nu
}
\right| .
\end{equation}

The product of two Levi-Civita symbols is defined in terms of the generalized Kronecker delta, however in the case of Minkowski spacetime this relationship has a sign change, since raising the indices in one of the symbols will produce an overall sign change. Therefore for 4D Minkowski spacetime follows the relationship $\epsilon_{\rho\sigma\mu\nu} \epsilon^{\alpha\beta\lambda\gamma} = - \delta_{\rho\sigma\mu\nu}^{\alpha\beta\lambda\gamma}$. Computing the determinant above is straightforward and yields,

\begin{equation}
\fl
\eqalign{ 
 \epsilon_{\rho\sigma\mu\nu} \epsilon^{\alpha\beta\lambda\gamma} =
- \delta_{\rho\sigma\mu\nu}^{\alpha\beta\lambda\gamma} =
\\
- \delta^\alpha_\rho  \delta^\beta_\sigma \delta^\lambda_\mu \delta^\gamma_\nu 
+\delta^\alpha_\rho \delta^\beta_\sigma \delta^\gamma_\mu \delta^\lambda_\nu
- \delta^\alpha_\rho\delta^\gamma_\sigma \delta^\beta_\mu  \delta^\lambda_\nu 
+\delta^\alpha_\rho \delta^\gamma_\sigma   \delta^\lambda_\mu \delta^\beta_\nu 
- \delta^\alpha_\rho \delta^\lambda_\sigma  \delta^\gamma_\mu \delta^\beta_\nu 
+\delta^\alpha_\rho \delta^\lambda_\sigma  \delta^\beta_\mu   \delta^\gamma_\nu 
\\
- \delta^\lambda_\rho   \delta^\alpha_\sigma  \delta^\beta_\mu  \delta^\gamma_\nu 
+ \delta^\lambda_\rho    \delta^\alpha_\sigma  \delta^\gamma_\mu  \delta^\beta_\nu 
- \delta^\lambda_\rho  \delta^\gamma_\sigma   \delta^\alpha_\mu  \delta^\beta_\nu 
+ \delta^\lambda_\rho  \delta^\gamma_\sigma   \delta^\beta_\mu  \delta^\alpha_\nu  
- \delta^\lambda_\rho  \delta^\beta_\sigma  \delta^\gamma_\mu  \delta^\alpha_\nu 
+ \delta^\lambda_\rho    \delta^\beta_\sigma  \delta^\alpha_\mu   \delta^\gamma_\nu  
\\
- \delta^\beta_\rho   \delta^\lambda_\sigma  \delta^\alpha_\mu \delta^\gamma_\nu 
+ \delta^\beta_\rho  \delta^\lambda_\sigma   \delta^\gamma_\mu  \delta^\alpha_\nu  
- \delta^\beta_\rho \delta^\gamma_\sigma  \delta^\lambda_\mu \delta^\alpha_\nu  
+ \delta^\beta_\rho \delta^\gamma_\sigma   \delta^\alpha_\mu   \delta^\lambda_\nu 
- \delta^\beta_\rho \delta^\alpha_\sigma  \delta^\gamma_\mu \delta^\lambda_\nu 
+ \delta^\beta_\rho  \delta^\alpha_\sigma  \delta^\lambda_\mu   \delta^\gamma_\nu 
\\
- \delta^\gamma_\rho  \delta^\beta_\sigma \delta^\alpha_\mu \delta^\lambda_\nu  
+ \delta^\gamma_\rho   \delta^\beta_\sigma \delta^\lambda_\mu   \delta^\alpha_\nu 
- \delta^\gamma_\rho \delta^\lambda_\sigma  \delta^\beta_\mu  \delta^\alpha_\nu 
+ \delta^\gamma_\rho \delta^\lambda_\sigma   \delta^\alpha_\mu   \delta^\beta_\nu 
- \delta^\gamma_\rho \delta^\alpha_\sigma  \delta^\lambda_\mu  \delta^\beta_\nu 
+ \delta^\gamma_\rho \delta^\alpha_\sigma  \delta^\beta_\mu   \delta^\lambda_\nu .
}   \label{genkron0}
\end{equation}

If two of these indices are contracted the above expression simplifies to,

\begin{equation}
\fl
 \epsilon_{\rho\sigma\mu\nu} \epsilon^{\alpha\beta\lambda\nu} =
- \delta_{\rho\sigma\mu\nu}^{\alpha\beta\lambda\nu} =
 \delta^\lambda_\rho \delta^\beta_\sigma \delta^\alpha_\mu 
 -   \delta^\beta_\rho \delta^\lambda_\sigma \delta^\alpha_\mu
 - \delta^\lambda_\rho \delta^\alpha_\sigma  \delta^\beta_\mu 
 + \delta^\alpha_\rho   \delta^\lambda_\sigma \delta^\beta_\mu 
 + \delta^\beta_\rho  \delta^\alpha_\sigma \delta^\lambda_\mu 
 -  \delta^\alpha_\rho  \delta^\beta_\sigma   \delta^\lambda_\mu .
\label{genkron1}
\end{equation}

If an additional two indices are contracted we are left with,

\begin{equation}
 \epsilon_{\rho\sigma\mu\nu} \epsilon^{\alpha\beta\mu\nu} =
- \delta_{\rho\sigma\mu\nu}^{\alpha\beta\mu\nu} =
- 2 (\delta^\alpha_\rho  \delta^\beta_\sigma  - \delta^\beta_\rho  \delta^\alpha_\sigma ) .
\label{genkron2}
\end{equation}

These expressions will be used to form identities between the dual and non-dual expressions. For example, the aforementioned relationship $M_1 = F_{\mu\nu} F^{\mu\nu} = -M_2 = - \mathcal{F}_{\mu\nu} \mathcal{F}^{\mu\nu}$ can be readily computed with $\mathcal{F}_{\mu\nu} =  \frac{1}{2} \epsilon_{\mu\nu\rho\sigma} {F}^{\rho\sigma}  $ as,

\begin{equation}
\hspace{-1cm}
- \mathcal{F}_{\mu\nu} \mathcal{F}^{\mu\nu} = - \frac{1}{4}  \epsilon_{\rho\sigma\mu\nu} \epsilon^{\alpha\beta\mu\nu} F_{\alpha\beta} F^{\rho\sigma} = \frac{1}{2}  (\delta^\alpha_\rho  \delta^\beta_\sigma  - \delta^\beta_\rho  \delta^\alpha_\sigma ) F_{\alpha\beta} F^{\rho\sigma} = F_{\mu\nu} F^{\mu\nu} . \label{maxscalariden}
\end{equation}

These generalized Kronecker deltas will be referred to throughout the article.

\subsection{Dualizing the non-homogenous half of electrodynamics}

The conventional Lagrangian for electrodynamic theory $\mathcal{L}_{MNH} = - \frac{1}{4} F_{\mu\nu} F^{\mu\nu}$ yields only half of Maxwell's equations in the Euler-Lagrange equation, namely the 4 non-homogenous equations known as the Gauss-Ampere laws in Equation (\ref{MaxEOMMNH}). Note that reference to anything associated to the non-homogenous half of Maxwell's equations from here forward will be denoted with subscript $MNH$ = Maxwell's non-homogenous for clarity. This Lagrangian is, of course, perfectly sound at deriving half of the theory from Noether's theorem, as shown in \cite{baker2019}. This non-homogenous equation of motion is dual to the equation of motion that represents the 4 homogenous equations, derived in Section \ref{maxhomogenous}.

Recall from Equation (\ref{maxscalariden}) the relationship $F_{\mu\nu} F^{\mu\nu} = -\mathcal{F}_{\mu\nu} \mathcal{F}^{\mu\nu}$. Since $\mathcal{L}_{MNH} = - \frac{1}{4} F_{\mu\nu} F^{\mu\nu}$ doesn't need to be changed, it can be expressed equivalently as $\mathcal{L}_{MNH} = \frac{1}{4} \mathcal{F}_{\mu\nu} \mathcal{F}^{\mu\nu}$ or $\mathcal{L}_{MNH} = \frac{1}{8} (\mathcal{F}_{\mu\nu} \mathcal{F}^{\mu\nu} - F_{\mu\nu} F^{\mu\nu})$. All of these options change sign under $F_{\mu\nu} \leftrightarrow \mathcal{F}_{\mu\nu}$, however the third is preferred since it is explicitly in dual form. Therefore the dual Lagrangian density is defined as the third option,

\begin{equation}
\mathcal{L}_{MNH} = \frac{1}{8} (\mathcal{F}_{\mu\nu} \mathcal{F}^{\mu\nu} - F_{\mu\nu} F^{\mu\nu}) . \label{MNHlag}
\end{equation}

The equations of motion that follows from substitution of this Lagrangian density into the Euler-Lagrange equation in Equation (\ref{genenergy}) are the 4 equations known as the non-homogenous half of Maxwell's equations (Gauss-Ampere laws). These equations are sourced by the 4-current $J^\mu$ that is coupled in to the conventional Lagrangian $\mathcal{L}_{MNH}$ via $A_\mu J^\mu$, hence the name non-homogenous. As in \cite{baker2019} we are only concerned with the free fields derived from the procedure, thus we will focus on the equations of motion $E^\rho_{MNH}$ that follows from $\mathcal{L}_{MNH}$ in Equation (\ref{MNHlag}) substituted into the Euler-Lagrange equation (Equation (\ref{genenergy})),

\begin{equation}
E^\rho_{MNH} = \partial_\sigma F^{\sigma\rho} . \label{MaxEOMMNH}
\end{equation}

Dual formulation of Maxwell's equations require the complete 8 equations, therefore the homogenous 4 are required, which are presented in Section \ref{maxhomogenous}. Finally, consider the energy-momentum tensor $T^{\mu\nu}_{MNH} = F^{\mu\alpha} F^\nu_{\ \alpha} - \frac{1}{4} \eta^{\mu\nu} F_{\alpha\beta} F^{\alpha\beta}$ derived from Noether's first theorem using Equations (\ref{genenergy}) and (\ref{MNHlag}). This can be elegantly dualized by deriving an identity relating the two terms above to the dual expression $\mathcal{F}^{\mu\alpha} \mathcal{F}^\nu_\alpha$ via equation (\ref{genkron2}) ,

\begin{equation}
\hspace{-0.5cm}
\fl
 \mathcal{F}^{\mu\alpha} \mathcal{F}^{\nu}_{\ \alpha} = \frac{1}{4} \eta^{\nu\omega}  \epsilon^{\rho\beta\mu\alpha} \epsilon_{\xi\sigma\omega\alpha} {F}_{\rho\beta}    {F}^{\xi\sigma}  = \frac{1}{4} \eta^{\nu\omega}  (-\delta^{\rho\beta\mu\alpha}_{\xi\sigma\omega\alpha}) {F}_{\rho\beta}    {F}^{\xi\sigma} = - \frac{1}{2} \eta^{\mu\nu}  F_{\alpha\beta} F^{\alpha\beta}
+   {F}^{\mu\alpha}  {F}^{\nu}_{\ \alpha} .   \label{MNHdualemtidentity}
\end{equation}

From this expression the term proportional to Minkowski can be re-expressed as $- \frac{1}{4}  \eta^{\mu\nu}  F_{\alpha\beta} F^{\alpha\beta} = \frac{1}{2} \mathcal{F}^{\mu\alpha} \mathcal{F}^{\nu}_{\ \alpha} -  \frac{1}{2}  {F}^{\mu\alpha}  {F}^{\nu}_{\ \alpha}$. Therefore the result for the dualized energy-momentum tensor is,

\begin{equation}
T^{\mu\nu}_{MNH} = \frac{1}{2} [{F}^{\mu\alpha}  {F}^{\nu}_{\ \alpha}   + \mathcal{F}^{\mu\alpha} \mathcal{F}^{\nu}_{\ \alpha}] . \label{TMNH}
\end{equation}

This equation is the dual form of the conventional energy-momentum tensor in electrodynamic theory and it is symmetric, conserved, gauge invariant, and has the additional explicit property of dual invariance under interchange $F_{\mu\nu} \leftrightarrow \mathcal{F}_{\mu\nu}$.

\subsection{Dualizing the homogenous half of electrodynamics \label{maxhomogenous}} 

In order to have the dual symmetry with the non-homogenous half of Maxwell's equations, the dual equation of motion is required, namely the 4 homogenous equations known as the Gauss-Faraday laws (Equation (\ref{mheomdual})) \cite{minkowski1909}. The only remaining invariant, $M_3 = {F}_{\mu\nu} \mathcal{F}^{\mu\nu}$, gives precisely this equation of motion in the Euler-Lagrange equation. This completes the dual symmetry of the equations of motion. Reference to anything associated to the homogenous half of Maxwell's equations from here forward will be denoted with subscript $MH$ = Maxwell's homogenous for clarity. The current section presents the dual form of the Lagrangian density $\mathcal{L}_{MH}$, equation of motion $E^\rho_{MH}$, and energy-momentum tensor $T^{\mu\nu}_{MH}$. Starting with the Lagrangian for the homogenous half of Maxwell's equations,

\begin{equation}
\mathcal{L}_{MH} = - \frac{1}{4} {F}_{\mu\nu} \mathcal{F}^{\mu\nu} .  \label{MHlag}
\end{equation}

Inserting this expression into the Euler-Lagrange equation in Equation (\ref{genenergy}), $\frac{\partial \mathcal{L}_{MH}}{\partial A_\rho} - \partial_\sigma \frac{\partial \mathcal{L}_{MH}}{\partial (\partial_\sigma A_\rho)} = \partial_\sigma \mathcal{F}^{\sigma\rho}$. Therefore Maxwell's homogenous equations are indeed the dual to the non-homogenous,

\begin{equation}
E^\rho_{MH} = \partial_\sigma \mathcal{F}^{\sigma\rho} .  \label{mheomdual}
\end{equation}

Note that using the definiton of the dual field strength tensor $\mathcal{F}_{\mu\nu} =  \frac{1}{2} \epsilon_{\mu\nu\rho\sigma} {F}^{\rho\sigma}  $, this equation can be re-expressed in terms of the Bianchi identity via $E^\rho_{MH} = \frac{1}{2} \epsilon^{\alpha\beta\sigma\rho} \partial_\sigma F_{\alpha\beta} = \frac{1}{6} \epsilon^{\alpha\beta\sigma\rho} (\partial_\sigma F_{\alpha\beta}+\partial_\beta F_{\sigma\alpha}+\partial_\alpha F_{\beta\sigma})$. Therefore an alternate form of the homogenous equations are in terms of the Bianchi identity as follows,

\begin{equation}
E^\rho_{MH} = \frac{1}{6} \epsilon^{\alpha\beta\sigma\rho} (\partial_\sigma F_{\alpha\beta}+\partial_\beta F_{\sigma\alpha}+\partial_\alpha F_{\beta\sigma}) = 0 . \label{mheombianchi}
\end{equation}

Commonly in the literature the homogenous equations are expressed as $\partial_\sigma F_{\alpha\beta}+\partial_\beta F_{\sigma\alpha}+\partial_\alpha F_{\beta\sigma} = 0$, due to the Bianchi identity representing the homogenous half of Maxwell's equations, this was the aforementioned idea of Einstein \cite{einstein1916}. The problem with this approach is that it cares not if half of Maxwell's theory is derived by the Euler-Lagrange equation; instead half of the theory is simply stated separately as a property of the field strength tensor. The dual formulation solves this problem elegantly.

Finally, an energy-momentum tensor can be derived from Noether's first theorem using Equations (\ref{genenergy}) and (\ref{MHlag}),

\begin{equation}
T^{\mu\nu}_{MH} = \mathcal{F}^{\mu\alpha} F^\nu_{\ \alpha} - \frac{1}{4} \eta^{\mu\nu} \mathcal{F}^{\alpha\beta} F_{\alpha\beta} ,
\end{equation}

which is also dually invariant under interchange $F_{\mu\nu} \leftrightarrow \mathcal{F}_{\mu\nu}$.\\

\section{Dual linearized Gauss-Bonnet gravity \label{s4}}

\subsection{Dualizing the Lagrangian density \label{s41}}

Now that the complete theory of electrodynamics has been expressed explicitly in the dual formulation, the completely gauge invariant linearized Gauss-Bonnet gravity model derived in [1] can be dualized. Recall that the Lagrangian was of the form 

\begin{equation}
\mathcal{L}_{LGB} = \frac{1}{4} (R_{\mu\nu\alpha\beta} R^{\mu\nu\alpha\beta} - 4 R_{\mu\nu} R^{\mu\nu} + R^2) , \label{LGBLAG1}
\end{equation}

where reference to anything associated to the linearized Gauss-Bonnet gravity model from here forward will have subscript $LGB$ = linearized Gauss-Bonnet for clarity. The scalars here are built from the linearized Riemann tensor $R^{\mu\nu\alpha\beta}$, linearized Ricci tensor $R^{\nu\beta}$, and linearized Ricci scalar $R$, respectively, 

\begin{equation}
R^{\mu\nu\alpha\beta} = \frac{1}{2} (  \partial^\mu \partial^\beta h^{\nu\alpha} + \partial^\nu \partial^\alpha h^{\mu\beta} -\partial^\mu \partial^\alpha h^{\nu\beta} - \partial^\nu \partial^\beta h^{\mu\alpha}),
\end{equation}

\begin{equation}
R^{\nu\beta} = \eta_{\mu\alpha} R^{\mu\nu\alpha\beta} = \frac{1}{2} (   \partial^\beta \partial^\alpha h_{\alpha}^{\nu}+\partial^\nu \partial^\alpha h_{\alpha}^{\beta} -\square h^{\nu\beta} - \partial^\nu \partial^\beta h ),
\end{equation}

\begin{equation}
R = \eta_{\nu\beta} R^{\nu\beta} =  \partial_\mu \partial_\nu h^{\mu\nu} - \square h .
\end{equation}

In the current section we will derive the dual form of the Lagrangian density $\mathcal{L}_{LGB}$, equation of motion $E^{\rho\sigma}_{LGB}$, and energy-momentum tensor $T^{\mu\nu}_{LGB}$. In order to dualize this Lagrangian, identities can be derived for the `doubly' dual linearized Riemann tensor $\mathcal{R}_{\mu\nu\alpha\beta}$, as well as the corresponding Ricci tensors and Ricci scalars.  For brevity the `doubly' dual Riemann tensor $\mathcal{R}_{\mu\nu\alpha\beta}$ in Equation (\ref{Riedoubdual}) will be referred to as the dual Riemann tensor; this is the dual tensor which we use in our article. From here the dual Ricci tensor $\mathcal{R}_{\mu\nu}$ and dual Ricci scalar $\mathcal{R}$ by contracting indices of the dual Riemann tensor,

\begin{equation}
\mathcal{R}_{\mu\nu} = \mathcal{R}_{\mu\alpha\nu\beta} \eta^{\alpha\beta} = \frac{1}{4} R^{\rho\sigma}_{\ \ \lambda\gamma} \epsilon_{\rho\sigma\mu\beta} \epsilon^{\lambda\gamma\alpha\beta} \eta_{\nu\alpha} , \label{Ricdoubdual}
\end{equation}

\begin{equation}
\mathcal{R} = \mathcal{R}_{\mu\nu} \eta^{\mu\nu} = \frac{1}{4} R^{\rho\sigma}_{\ \ \lambda\gamma} \epsilon_{\rho\sigma\alpha\beta} \epsilon^{\lambda\gamma\alpha\beta} . \label{Ricscalrdoubdual}
\end{equation}

We note that there is also the `simply' dual ${\bf{R}}_{\alpha\beta\mu\nu}$ in Equation (\ref{Riesingdual}) that dualizes only one of the antisymmetric pairs of the Riemann tensor. Scalars from this expression, such as the Lanczos $K_1 = {\bf{R}}^{\alpha\beta\mu\nu} R_{\alpha\beta\mu\nu}$, are redundant to what can be found with $\mathcal{R}_{\mu\nu\alpha\beta}$. Furthermore, the Ricci tensor and Ricci scalar duals from ${\bf{R}}_{\alpha\beta\mu\nu}$ are identically zero (${\bf{R}}_{\mu\nu} = 0$, ${\bf{R} = 0}$), due to the first Bianchi identity. 

Using the dual Riemann expressions in Equations (\ref{Riedoubdual}), (\ref{Ricdoubdual}) and (\ref{Ricscalrdoubdual}) the following identities can be derived by using the generalized Kronecker delta in Equations (\ref{genkron1}) and (\ref{genkron2}) on the combinations $\mathcal{R}_{\mu\nu\alpha\beta} \mathcal{R}^{\mu\nu\alpha\beta}$, $\mathcal{R}_{\mu\nu} \mathcal{R}^{\mu\nu}$ and $\mathcal{R}^2$,

\begin{equation}
\hspace{-1cm}
\fl
\mathcal{R}_{\mu\nu\alpha\beta} \mathcal{R}^{\mu\nu\alpha\beta} 
= \frac{1}{16} R_{\alpha\beta}^{\ \ \omega\tau} \epsilon^{\alpha\beta\mu\nu} \epsilon_{\omega\tau\lambda\gamma} R^{\rho\sigma}_{\ \ \theta \phi} \epsilon_{\rho\sigma\mu\nu} \epsilon^{\theta\phi\lambda\gamma} 
= \frac{1}{16} R_{\alpha\beta}^{\ \ \omega\tau} R^{\rho\sigma}_{\ \ \theta \phi} (-\delta^{\alpha\beta\mu\nu}_{\rho\sigma\mu\nu})(- \delta_{\omega\tau\lambda\gamma}^{\theta\phi\lambda\gamma}) 
= {R}_{\mu\nu\alpha\beta} {R}^{\mu\nu\alpha\beta} , \label{rieriedauliden}
\end{equation}

\begin{equation}
\hspace{-1cm}
\fl
\mathcal{R}_{\mu\nu} \mathcal{R}^{\mu\nu} 
= \frac{1}{16} R^{\rho\sigma}_{\ \ \theta \gamma} \epsilon_{\rho\sigma\mu\beta} \epsilon^{\theta\gamma\alpha\beta} \eta_{\alpha\lambda} R_{\omega\tau}^{\ \ \xi\chi}  \epsilon^{\omega\tau\mu\phi} \epsilon_{\xi\chi\delta\phi} \eta^{\delta\lambda}
= \frac{1}{16} R^{\rho\sigma}_{\ \ \theta \gamma} R_{\omega\tau}^{\ \ \xi\chi} (-\delta_{\rho\sigma\beta\mu}^{\omega\tau\phi\mu})( -\delta^{\theta\gamma\beta\delta}_{\xi\chi\phi\delta}) 
= {R}_{\mu\nu} {R}^{\mu\nu} ,
\end{equation}

\begin{equation}
\hspace{-1cm}
\fl
\mathcal{R}^2
= \frac{1}{16} R^{\rho\sigma}_{\ \ \lambda\gamma} \epsilon_{\rho\sigma\alpha\beta} \epsilon^{\lambda\gamma\alpha\beta} R^{\mu\nu}_{\ \ \omega\tau} \epsilon_{\mu\nu\theta\phi} \epsilon^{\omega\tau\theta\phi}
= \frac{1}{16} R^{\rho\sigma}_{\ \ \lambda\gamma} R^{\mu\nu}_{\ \ \omega\tau} (-\delta_{\rho\sigma\alpha\beta}^{\lambda\gamma\alpha\beta})(-\delta_{\mu\nu\theta\phi}^{\omega\tau\theta\phi}) .
= R^2
\end{equation}

Similar to electrodynamics, each of the dual scalars can be expressed in terms of their corresponding original non-dual scalars. An equivalent Lagrangian for the linearized Gauss-Bonnet model $\mathcal{L} = \frac{1}{4} (R_{\mu\nu\alpha\beta} R^{\mu\nu\alpha\beta} - 4 R_{\mu\nu} R^{\mu\nu} + R^2)$ can therefore be expressed as $\mathcal{L} = \frac{1}{4} (\mathcal{R}_{\mu\nu\alpha\beta} \mathcal{R}^{\mu\nu\alpha\beta}  - 4 \mathcal{R}_{\mu\nu} \mathcal{R}^{\mu\nu}  + \mathcal{R}^2)$. Writing this in dually symmetric form, similar to $\mathcal{L}_{MNH}$, the resulting Lagrangian density is,

\begin{equation}
\fl
\mathcal{L}_{LGB} = \frac{1}{8} (\mathcal{R}_{\mu\nu\alpha\beta} \mathcal{R}^{\mu\nu\alpha\beta}  - 4 \mathcal{R}_{\mu\nu} \mathcal{R}^{\mu\nu}  + \mathcal{R}^2 + {R}_{\mu\nu\alpha\beta} {R}^{\mu\nu\alpha\beta}  - 4 {R}_{\mu\nu} {R}^{\mu\nu}  + {R}^2) . \label{LLGBMNH}
\end{equation}

This presentation would suggest analogy to the $MNH$ equations. The above expression is invariant under interchange ${R}_{\mu\nu\alpha\beta} \leftrightarrow \mathcal{R}_{\mu\nu\alpha\beta}$, ${R}_{\mu\nu} \leftrightarrow \mathcal{R}_{\mu\nu}$ and ${R} \leftrightarrow \mathcal{R}$. However, recalling the invariant presented by Lanczos $K_2 = \mathcal{R}^{\alpha\beta\mu\nu} R_{\alpha\beta\mu\nu}$, who noticed the relationship $K_2 \propto I_3 - 4 I_1 + I_2$, indeed deriving the identity for $K_2$ yields,

\begin{equation}
\fl
\mathcal{R}_{\mu\nu\alpha\beta} {R}^{\mu\nu\alpha\beta} 
= \frac{1}{4} R^{\rho\sigma}_{\ \ \lambda \gamma} R_{\alpha\beta}^{\ \ \mu\nu} (-\delta_{\rho\sigma\mu\nu}^{\alpha\beta\lambda\gamma}) = - {R}_{\mu\nu\alpha\beta} {R}^{\mu\nu\alpha\beta}  + 4 {R}_{\mu\nu} {R}^{\mu\nu}  - {R}^2 . \label{gbldualiden}
\end{equation}

Using this identity the Lagrangian $\mathcal{L}_{LGB}$ can also be expressed as,

\begin{equation}
\mathcal{L}_{LGB} = - \frac{1}{4} \mathcal{R}_{\mu\nu\alpha\beta} {R}^{\mu\nu\alpha\beta} .  \label{LLGBMH}
\end{equation}

This presentation seems to indicate an analogy with the $MH$ equations. While the Lagrangian $\mathcal{L}_{LGB}$ can be expressed in dual form analogous to both halves of Maxwell's theory, this discrepancy will be clearly avoided for the equation of motion $E^{\rho\sigma}_{LGB}$ in the following sections, which corresponds to the second Bianchi identity as in the $MH$ case. First the dualization of the energy-momentum tensor will be performed.

\subsection{Dualizing the energy-momentum tensor}

The energy-momentum tensor for the linearized Gauss-Bonnet gravity model, a well known expression given in 
Equation (\ref{gaussemt}), was derived from Noether's theorem in \cite{baker2019}. To dualize this expression, a series of identities can be derived relating the terms in the energy-momentum tensor to the corresponding dual terms, as in the case of Equation (\ref{MNHdualemtidentity}). For the four terms in Equation (\ref{gaussemt}) not proportional to Minkowski ($-  R^{\omega\rho\lambda\sigma} R^\nu_{\ \rho\lambda\sigma} + 2  R_{\rho\sigma} R^{\omega \rho \nu \sigma} + 2  R^{\omega\lambda} R^\nu_{\ \lambda} -  R R^{\omega\nu}$), the identities between dual and non-dual are, from Equations (\ref{Riedoubdual}), (\ref{Ricdoubdual}) and (\ref{Ricscalrdoubdual}) and using Equations (\ref{genkron1}) and (\ref{genkron2}),

\begin{equation}
\hspace{-1cm}
\fl
\mathcal{R}^{\omega\rho\lambda\sigma} \mathcal{R}^{\nu}_{\ \rho\lambda\sigma} 
= \frac{1}{16} \eta^{\nu\gamma} R_{\alpha\beta}^{\ \ \mu\xi} R_{\chi\delta}^{\ \ \theta\phi} (-\delta_{\mu\xi\lambda\sigma}^{\chi\delta\lambda\sigma}) (-\delta_{\theta\phi\gamma\rho}^{\alpha\beta\omega\rho})
= -  R^{\omega\rho\lambda\sigma} R^\nu_{\ \rho\lambda\sigma} + \frac{1}{2} \eta^{\omega\nu} {R}_{\mu\gamma\alpha\beta} {R}^{\mu\gamma\alpha\beta} , \label{drriden}
\end{equation}

\begin{equation}
\hspace{-1cm}
\fl
\mathcal{R}_{\rho\sigma} \mathcal{R}^{\omega\rho\nu\sigma} 
= \frac{1}{16} \eta^{\nu\gamma} R^{\mu\gamma}_{\ \ \beta\delta} R_{\theta\phi}^{\ \ \chi\xi} (-\delta_{\mu\gamma\alpha\rho}^{\theta\phi\omega\rho}) (-\delta_{\chi\xi\lambda\sigma}^{\beta\delta\alpha\sigma})
= R_{\rho\sigma} R^{\omega \rho \nu \sigma} -  R^{\omega\rho\lambda\sigma} R^\nu_{\ \rho\lambda\sigma} + \frac{1}{4} \eta^{\omega\nu} {R}_{\mu\gamma\alpha\beta} {R}^{\mu\gamma\alpha\beta} ,
\end{equation}

\begin{equation}
\hspace{-1cm}
\fl
\mathcal{R}^{\omega\lambda} \mathcal{R}^\nu_{\ \lambda}
= \frac{1}{16} \eta^{\gamma\lambda} R_{\mu\tau}^{\ \ \rho\sigma} R_{\theta\phi}^{\ \ \chi\xi} (-\delta_{\rho\sigma\gamma\alpha}^{\mu\tau\omega\alpha}) (-\delta_{\chi\xi\lambda\beta}^{\theta\phi\nu\beta})
= R^{\omega\lambda} R^\nu_{\ \lambda} -  R R^{\omega\nu} + \frac{1}{4} \eta^{\omega\nu} R^2 ,
\end{equation}

\begin{equation}
\hspace{-1cm}
\fl
\mathcal{R} \mathcal{R}^{\omega\nu}
= \frac{1}{16} \eta^{\nu\gamma} R^{\mu\tau}_{\ \ \rho\sigma} R_{\theta\phi}^{\ \ \chi\xi} (-\delta_{\mu\tau\alpha\beta}^{\rho\sigma\alpha\beta}) (-\delta_{\chi\xi\gamma\lambda}^{\theta\phi\omega\lambda})
= - {R} {R}^{\omega\nu} + \frac{1}{2} \eta^{\omega\nu} R^2 . \label{drriden2}
\end{equation}

Combining these 4 terms in the manner they appear in the energy-momentum tensor yields an interesting identity,

\begin{equation}
\hspace{-1cm}
\fl
- \mathcal{R}^{\omega\rho\lambda\sigma} \mathcal{R}^{\nu}_{\ \rho\lambda\sigma} 
+ 2 \mathcal{R}_{\rho\sigma} \mathcal{R}^{\omega\rho\nu\sigma} 
+ 2 \mathcal{R}^{\omega\lambda} \mathcal{R}^\nu_{\ \lambda}
- \mathcal{R} \mathcal{R}^{\omega\nu}
=
-  R^{\omega\rho\lambda\sigma} R^\nu_{\ \rho\lambda\sigma} 
+ 2  R_{\rho\sigma} R^{\omega \rho \nu \sigma} 
+ 2  R^{\omega\lambda} R^\nu_{\ \lambda} 
-  R R^{\omega\nu} ,
\end{equation}

as the particular coefficients of the energy-momentum tensor cancel all of the second and third terms in Equations (\ref{drriden}) - (\ref{drriden2}). Since the term proportional to Minkowski can also be re-expressed $\mathcal{R}_{\mu\nu\alpha\beta} \mathcal{R}^{\mu\nu\alpha\beta}  - 4 \mathcal{R}_{\mu\nu} \mathcal{R}^{\mu\nu}  + \mathcal{R}^2 = {R}_{\mu\nu\alpha\beta} {R}^{\mu\nu\alpha\beta}  - 4 {R}_{\mu\nu} {R}^{\mu\nu}  + {R}^2$ as shown in Section \ref{s41}, the energy-momentum tensor in Equation (\ref{gaussemt}) can be expressed as $T^{\omega\nu}_{LGB} = - \mathcal{R}^{\omega\rho\lambda\sigma} \mathcal{R}^{\nu}_{\ \rho\lambda\sigma} + 2 \mathcal{R}_{\rho\sigma} \mathcal{R}^{\omega\rho\nu\sigma} + 2 \mathcal{R}^{\omega\lambda} \mathcal{R}^\nu_{\ \lambda} - \mathcal{R} \mathcal{R}^{\omega\nu} + \frac{1}{4} \eta^{\omega\nu} (\mathcal{R}_{\mu\gamma\alpha\beta} \mathcal{R}^{\mu\gamma\alpha\beta}  - 4 \mathcal{R}_{\mu\gamma} \mathcal{R}^{\mu\gamma}  + \mathcal{R}^2) $. The goal of this section is to write the energy-momentum tensor in dually invariant form, thus a third equivalent representation based on the dual and non-dual is,

\begin{equation}
\fl
\eqalign{ 
T^{\omega\nu}_{LGB} = -  \frac{1}{2} R^{\omega\rho\lambda\sigma} R^\nu_{\ \rho\lambda\sigma} +   R_{\rho\sigma} R^{\omega \rho \nu \sigma} +   R^{\omega\lambda} R^\nu_{\ \lambda} - \frac{1}{2}  R R^{\omega\nu}
\\
- \frac{1}{2} \mathcal{R}^{\omega\rho\lambda\sigma} \mathcal{R}^{\nu}_{\ \rho\lambda\sigma} 
+  \mathcal{R}_{\rho\sigma} \mathcal{R}^{\omega\rho\nu\sigma} +  \mathcal{R}^{\omega\lambda} \mathcal{R}^\nu_{\ \lambda}
- \frac{1}{2}  \mathcal{R} \mathcal{R}^{\omega\nu} 
\\
+ \frac{1}{8} \eta^{\omega\nu} ({R}_{\mu\gamma\alpha\beta} {R}^{\mu\gamma\alpha\beta}  - 4 {R}_{\mu\gamma} {R}^{\mu\gamma}  + {R}^2)
+ \frac{1}{8} \eta^{\omega\nu} (\mathcal{R}_{\mu\gamma\alpha\beta} \mathcal{R}^{\mu\gamma\alpha\beta}  - 4 \mathcal{R}_{\mu\gamma} \mathcal{R}^{\mu\gamma}  + \mathcal{R}^2) .
}
\end{equation}

This presentation of the energy-momentum tensor is explicitly invariant under interchange ${R}_{\mu\nu\alpha\beta} \leftrightarrow \mathcal{R}_{\mu\nu\alpha\beta}$, ${R}_{\mu\nu} \leftrightarrow \mathcal{R}_{\mu\nu}$ and ${R} \leftrightarrow \mathcal{R}$. This form, similar to the first $\mathcal{L}_{LGB}$ derived in Equation (\ref{LLGBMNH}), is analogous to what is found for $T^{\omega\nu}_{MNH}$ in the $MNH$ half of electrodynamics. Another dually invariant and equivalent expression can be considered by deriving an identity between the dual and non-dual tensors $\mathcal{R}^{\omega\alpha\beta\lambda} {R}^{\nu}_{\ \alpha\beta\lambda} $,

\begin{equation}
\hspace{-1cm}
\fl
\mathcal{R}^{\omega\alpha\beta\lambda} {R}^{\nu}_{\ \alpha\beta\lambda} 
= \frac{1}{4} \eta^{\theta\nu} R_{\alpha\beta}^{\ \ \rho\sigma} R_{\theta\gamma}^{\ \ \mu\phi} (-\delta_{\rho\sigma\mu\phi}^{\alpha\beta\omega\gamma}) 
= -  R^{\omega\rho\lambda\sigma} R^\nu_{\ \rho\lambda\sigma} + 2  R_{\rho\sigma} R^{\omega \rho \nu \sigma} + 2  R^{\omega\lambda} R^\nu_{\ \lambda} -  R R^{\omega\nu} .
\end{equation}

This is exactly the non-Minkowski part of Equation (\ref{gaussemt})! In addition, the Minkowski part in Equation (\ref{gaussemt}) can be re-expressed using $\mathcal{R}^{\mu\nu\alpha\beta} {R}_{\mu\nu\alpha\beta} = - {R}_{\mu\nu\alpha\beta} {R}^{\mu\nu\alpha\beta}  + 4 {R}_{\mu\nu} {R}^{\mu\nu}  - {R}^2$ from Equation (\ref{gbldualiden}), yielding a compact expression for the energy-momentum tensor of Gauss-Bonnet gravity in dual form,

\begin{equation}
T^{\omega\nu}_{LGB} =
\mathcal{R}^{\omega\alpha\beta\lambda} {R}^{\nu}_{\ \alpha\beta\lambda} 
- \frac{1}{4} \eta^{\omega\nu} \mathcal{R}^{\mu\gamma\alpha\beta} {R}_{\mu\gamma\alpha\beta} . \label{lgbdualenmo}
\end{equation}

In this presentation the energy-momentum tensor, similar to the second $\mathcal{L}_{LGB}$ derived in Equation (\ref{LLGBMH}), is analogous to what is found for $T^{\omega\nu}_{MH}$ in the $MH$ half of electrodynamics. Both the Lagrangian $\mathcal{L}_{LGB}$ and energy-momentum tensor $T^{\omega\nu}_{LGB}$ can be expressed in dual form analogous to both halves on Maxwell's equations. It appears however that the homogenous half $MH$ is truly analogous given compactness of these Equations (\ref{LLGBMH}) and (\ref{lgbdualenmo}), and the second Bianchi identity equation of motion. This will be evidenced by the equation of motion $E^{\rho\sigma}_{LGB}$ in dual form, which is the topic of the following section.

\subsection{Dualizing the equation of motion}

Conventional wisdom states that the {\it{`In D = 4 the Gauss-Bonnet invariant is a total derivative, and hence does not contribute
to the gravitational dynamics'}} \cite{glavan2020}, and more specifically to the equation of motion, {\it{`In the four-dimensional spacetime, the Gauss-Bonnet term does not contribute to the field equations since it becomes a total derivative'}} \cite{maeda2008}. This sentiment implies that there is simply nothing in the equation of motion following from the Gauss-Bonnet Lagrangian. A closer analysis shows that this is not the case. Differentiating the Gauss-Bonnet Lagrangian,

\begin{equation}
 \frac{\partial \mathcal{L}_{LGB}}{\partial (\partial_\omega \partial_\lambda h_{\rho\sigma})} =
\frac{1}{8} R_{\mu\nu\alpha\beta} [\epsilon^{\lambda\sigma\mu\nu} \epsilon^{\omega\rho\alpha\beta} + \epsilon^{\lambda\rho\mu\nu} \epsilon^{\omega\sigma\alpha\beta}] .
\end{equation}

Substituting this into the Euler-Lagrange equation in Equation (\ref{genenergy}), using the $\partial_\omega \partial_\lambda$ and $R_{\mu\nu\alpha\beta}$ symmetries, and reintroducing the dual $\mathcal{R}^{\omega\rho\lambda\sigma}$ in Equation (\ref{Riedoubdual}),

\begin{equation}
E^{\rho\sigma}_{LGB} =
\partial_\omega \partial_\lambda \frac{\partial \mathcal{L}_{LGB}}{\partial (\partial_\omega \partial_\lambda h_{\rho\sigma})} =
\frac{1}{4} \partial_\omega \partial_\lambda R_{\mu\nu\alpha\beta} \epsilon^{\lambda\mu\nu\sigma} \epsilon^{\omega\alpha\beta\rho} 
= \partial_\omega \partial_\lambda \mathcal{R}^{\omega\rho\lambda\sigma} .
\end{equation}

The equation of motion for linearized Gauss-Bonnet gravity is based on the second order divergence of the dual Riemann tensor, analogous to how equation (\ref{mheomdual}) is the divergence of the dual tensor for the homogenous half of electrodynamics. Similarly, using the Levi-Civita symbol, this can be re-expressed as the second Bianchi identity via $\partial_\lambda R_{\mu\nu\alpha\beta} \epsilon^{\lambda\mu\nu\sigma}  = \frac{1}{3} \epsilon^{\lambda\mu\nu\sigma} (\partial_\lambda R_{\mu\nu\alpha\beta} + \partial_\mu R_{\nu\lambda\alpha\beta} + \partial_\nu R_{\lambda\mu\alpha\beta}) = 0$. Therefore the equation of motion for the linearized Gauss-Bonnet model can be expressed as the second Bianchi identity as,

\begin{equation}
E^{\rho\sigma}_{LGB} =
\frac{1}{12} \epsilon^{\lambda\mu\nu\sigma} \epsilon^{\omega\alpha\beta\rho}  \partial_\omega (\partial_\lambda R_{\mu\nu\alpha\beta} + \partial_\mu R_{\nu\lambda\alpha\beta} + \partial_\nu R_{\lambda\mu\alpha\beta})
= 0 .
\end{equation}

The linearized Gauss-Bonnet model can therefore be completely expressed in analogous dually invariant form to the homogenous half of electrodynamics ($MH$) in Section \ref{maxhomogenous}. The equation of motion for both of these models is the second Bianchi identity. This raises a point, perhaps of fundamental significance; if the Bianchi identity which represents half of Maxwell's equations is considered a fundamental equation of motion to electrodynamics, should the second Bianchi identity of the Riemann tensor be thought of as part of the fundamental set of equations for the Gauss-Bonnet theories of gravity, or more generally, metric theories of gravity? Such views have been considered in the literature in the past \cite{kling2005}, but are not often included in the set of fundamental equations of motion as in the case of classical electrodynamics.

\section{Completing the dual linearized gravity model \label{s5}}

The linearized Gauss-Bonnet gravity model has been expressed in dual form, with $\mathcal{L}_{LGB}$ and $T^{\mu\nu}_{LGB}$ independently dual invariant analogous to the homogenous half of electrodynamics ($MH$) in Section \ref{maxhomogenous}. This analogy was further emphasized by the second Bianchi identity being the equation of motion for the model. One major issue arises here, however, in the fact that the equation of motion itself does not have a dual counterpart which can be found under interchange of the Riemann tensor ${R}_{\mu\nu\alpha\beta} \leftrightarrow \mathcal{R}_{\mu\nu\alpha\beta}$. This is a major issue for three reasons: (i) in order to introduce a dual equation of motion, another internally dual invariant must be introduced to the Lagrangian as in the $MNH$ case, (ii) the possible Lagrangian densities ($\mathcal{L} = \tilde{a} R_{\mu\nu\alpha\beta} R^{\mu\nu\alpha\beta} + \tilde{b} R_{\mu\nu} R^{\mu\nu} + \tilde{c} R^2$) are constrained by the procedure in \cite{baker2019}, and (iii) the procedure in \cite{baker2019} showed that the Gauss-Bonnet energy-momentum tensor was the unique gauge invariant, symmetric and trace-free expression derived from Noether's theorem for these possible Lagrangian densities.

Problems (i) and (ii) can be easily remedied by noticing the dual equation of motion to $E^{\rho\sigma}_{LGB}$ is trivially of the form $\partial_\omega \partial_\lambda {R}^{\omega\rho\lambda\sigma}$, which follows from one of the constrained invariants ${R}_{\mu\nu\alpha\beta} {R}^{\mu\nu\alpha\beta}$, contraction of the linearized Riemann tensors. The model built from this scalar alone will now be explored. Reference to anything associated to the dual form of model derived from the linearized Riemann-Riemann scalar from here forward will have subscript $LRR$ = linearized Riemann-Riemann for clarity. Therefore the current section explores the dual form of the Lagrangian density $\mathcal{L}_{LRR}$, equation of motion $E^{\rho\sigma}_{LRR}$, and energy-momentum tensor $T^{\mu\nu}_{LRR}$. Problem (iii) is significantly less trivial and will be discussed in detail.

\subsection{Dualizing the Lagrangian and equation of motion}

In order to have the equation of motion dual to $E^{\rho\sigma}_{LGB}$, a Lagrangian of the form ${R}_{\mu\nu\alpha\beta} {R}^{\mu\nu\alpha\beta}$ will be considered, namely $\mathcal{L}_{LRR} = - \frac{1}{4} {R}_{\mu\nu\alpha\beta} {R}^{\mu\nu\alpha\beta}$. From Equation (\ref{rieriedauliden}), the dualization is trivial, since ${R}_{\mu\nu\alpha\beta} {R}^{\mu\nu\alpha\beta} = \mathcal{R}_{\mu\nu\alpha\beta} \mathcal{R}^{\mu\nu\alpha\beta} $. Therefore it can equivalently be expressed as $\mathcal{L}_{LRR} = - \frac{1}{4} \mathcal{R}_{\mu\nu\alpha\beta} \mathcal{R}^{\mu\nu\alpha\beta} $, and in dually symmetric form as,

\begin{equation}
\mathcal{L}_{LRR} = - \frac{1}{8} ( {R}_{\mu\nu\alpha\beta} {R}^{\mu\nu\alpha\beta} + \mathcal{R}_{\mu\nu\alpha\beta} \mathcal{R}^{\mu\nu\alpha\beta} ) .
\end{equation}

This dual Lagrangian is strikingly similar to that of the non-homogenous ($MNH$) half of electrodynamics. Differentiating this expression yields $\frac{\partial \mathcal{L}_{LRR}}{\partial (\partial_\omega \partial_\lambda h_{\rho\sigma})} =  \frac{1}{2} [R^{\omega\rho\lambda\sigma} +  R^{\lambda\rho\omega\sigma}] $. The Euler-Lagrange equation of motion from Equation (\ref{genenergy}) is therefore,

\begin{equation}
E^{\rho\sigma}_{LRR} =
\partial_\omega \partial_\lambda \frac{\partial \mathcal{L}_{LRR}}{\partial (\partial_\omega \partial_\lambda h_{\rho\sigma})} =
 \partial_\omega \partial_\lambda {R}^{\omega\rho\lambda\sigma} ,
\end{equation}

which is indeed the expression dual to $E^{\rho\sigma}_{LGB}$. The $LRR$ model therefore has an internally dual symmetric Lagrangian, and an equation of motion dual to that of linearized Gauss-Bonnet gravity; both which are analogous to the non-homogenous half of electrodynamics. Problem (iii) is now to derive the energy-momentum tensor, which is not trivially gauge invariant as $T^{\mu\nu}_{LGB}$ is in \cite{baker2019}.

\subsection{Dualizing the energy-momentum tensor}

In [1] the linearized Gauss-Bonnet gravity model was the unique model derived from the procedure for $N = M = 2$ and had a gauge invariant energy-momentum tensor. To understand why this is, we must consider the conserved current from Noether's first theorem in Equation (\ref{genenergy}) for a Lagrangian density of the form $\partial \partial h \partial \partial h$,

  \begin{equation}
   \partial_\omega \left[ 
  \frac{\partial \mathcal{L}}{\partial (\partial_\omega \partial_\lambda h_{\rho\sigma})} \partial_\lambda \delta h_{\rho\sigma} 
+ \eta^{\omega\nu} \mathcal{L} \delta x_\nu 
 - \left( \partial_\lambda \frac{\partial \mathcal{L}}{\partial (\partial_\omega \partial_\lambda h_{\rho\sigma})} \right) \delta h_{\rho\sigma} \right] = 0. \label{ourgenen}
\end{equation}

The third term above is responsible for the lack of gauge invariance in the model, since the transformation $\delta h_{\rho\sigma} = - 2 \Gamma^\nu_{\ \rho \sigma} \delta x_\nu$ \cite{baker2019,besselhagen1921,jackiw1978}, where $\Gamma^\nu_{\ \rho \sigma} = \frac{1}{2}(\partial^\nu h_{\rho\sigma} - \partial_\rho h^\nu_\sigma - \partial_\sigma h^\nu_\rho)$, is not gauge invariant. This is in essence the same reason for the no-go result that spin-2 linearized gravity cannot have a gauge invariant energy-momentum tensor \cite{baker2019,padmanabhan2008,magnano2002,dewit1980,deser2010}, at least second order derivatives are needed. Only for the linearized Gauss-Bonnet Lagrangian density does the particular combination of invariants kill this term, resulting in a gauge invariant expression. The first term is gauge invariant because of $\partial_\lambda \delta h_{\rho\sigma} $ yielding the linearized Riemann tensor via $R^{\nu}_{\ \rho\sigma\lambda} = \partial_\lambda \Gamma^\nu_{\rho\sigma} - \partial_\sigma \Gamma^\nu_{\rho\lambda}$ which is independently gauge invariant. To show this explicitly, deriving the energy-momentum tensor from $\mathcal{L}_{LRR}$ in Equation (\ref{ourgenen}),

  \begin{equation}
\fl
T^{\omega\nu}_{LRR} =
R^{\omega\rho\lambda\sigma} R^\nu_{\ \rho\lambda\sigma} -\frac{1}{4}  \eta^{\omega\nu} {R}_{\mu\gamma\alpha\beta} {R}^{\mu\gamma\alpha\beta} - 2  \partial_\lambda  R^{\omega\rho\lambda\sigma}  \Gamma^\nu_{\ \rho \sigma} .
\end{equation}

This expression is not dually symmetric, not gauge invariant and not trace-free. It is merely conserved on-shell via $E^{\rho\sigma}_{LRR}$. This feature is expected since it is related to the main result of \cite{baker2019}, however it greatly hampers the development of a complete dual formulation. There are only two possible remedies to this problem. One is to integrate by parts the third term in the energy-momentum tensor $T^{\omega\nu}_{LRR}$, since one term will combine with the first time in $T^{\omega\nu}_{LRR}$ and the other term will be a second order term of the form $\partial_\omega  \partial_\lambda  (R^{\omega\rho\lambda\sigma}  \Gamma^\nu_{\ \rho \sigma})$. What remains under $\partial_\omega $ is $2 R^{\omega\rho\lambda\sigma} R^\nu_{\ \rho\lambda\sigma} -\frac{1}{4}  \eta^{\omega\nu} {R}_{\mu\gamma\alpha\beta} {R}^{\mu\gamma\alpha\beta}$ which is indeed gauge invariant and symmetric, but is neither trace-free nor conserved. Additionally the second order term does not trivially vanish because the symmetries $\omega\lambda$ and $\rho\sigma$ are unable to yield the first Bianchi identity. 

The second, and more reasonable solution, is to integrate by parts the equation of motion. This is possible because Noether's first theorem is used to derive a complete identity given in Equation ($\ref{genenergy}$), it is not simply a method of deriving equations of motion and conservation laws separately. For $LRR$ this yields $- 2 E^{\rho\sigma}_{LRR} \Gamma^\nu_{\ \rho \sigma} + \partial_\omega T^{\omega\nu}_{LRR} = 0$, which expands to,

\begin{equation}
\fl
2 \partial_\omega \partial_\lambda {R}^{\omega\rho\lambda\sigma} \Gamma^\nu_{\ \rho \sigma} + \partial_\omega (R^{\omega\rho\lambda\sigma} R^\nu_{\ \rho\lambda\sigma} -\frac{1}{4}  \eta^{\omega\nu} {R}_{\mu\gamma\alpha\beta} {R}^{\mu\gamma\alpha\beta} - 2  \partial_\lambda  R^{\omega\rho\lambda\sigma}  \Gamma^\nu_{\ \rho \sigma}) = 0 .
\end{equation}

Integration by parts of the first term (equation of motion) via $ 2  \partial_\omega \partial_\lambda {R}^{\omega\rho\lambda\sigma} \Gamma^\nu_{\ \rho \sigma}  = 2  \partial_\omega [\partial_\lambda {R}^{\omega\rho\lambda\sigma} \Gamma^\nu_{\ \rho \sigma}] -  2 \partial_\lambda {R}^{\omega\rho\lambda\sigma} \partial_\omega \Gamma^\nu_{\ \rho \sigma}$ exactly kills the third term in the energy-momentum expression. What is left in place of the equation of motion is $  -  2 \partial_\lambda {R}^{\omega\rho\lambda\sigma} \partial_\omega \Gamma^\nu_{\ \rho \sigma} = -   (\partial_\lambda {R}^{\omega\rho\lambda\sigma}) R^{\nu}_{\ \rho\omega\sigma}$, therefore we have the relationship between this expression and the total divergence from Noether's first theorem as,

\begin{equation}
(\partial_\lambda {R}^{\omega\rho\lambda\sigma}) R^{\nu}_{\ \rho\omega\sigma}  = \partial_\omega (R^{\omega\rho\lambda\sigma} R^\nu_{\ \rho\lambda\sigma} -\frac{1}{4}  \eta^{\omega\nu} {R}_{\mu\gamma\alpha\beta} {R}^{\mu\gamma\alpha\beta} ) .
\end{equation}

What is left in the divergence, is a precisely gauge invariant, symmetric and trace-free energy-momentum tensor! This expression is conserved on-shell via the equation $\bar{E}^{\rho\lambda\sigma}_{DRR} = \partial_\lambda {R}^{\omega\rho\lambda\sigma}$. This equation of motion can be expressed in dual form with the Gauss-Bonnet equation of motion, since the Gauss-Bonnet equation of motion is the second Bianchi identity which requires only one derivative of the dual Riemann tensor $\partial_\lambda \mathcal{R}^{\omega\rho\lambda\sigma}$, therefore not impacting the Lagrangian $\mathcal{L}_{LGB}$ or energy-momentum tensor $T^{\omega\nu}_{LGB}$ derived in \cite{baker2019}.  Furthermore, this energy-momentum tensor can be expressed in dually symmetric form analogous to the non-homogenous half of electrodynamics.

Since this presentation differs from that to this point (limiting the equation of motion to a single divergence of the field strength tensors), the dual form of the Riemann-Riemann model will now be referred to with subscript $DRR$ for clarity. The associated Gauss-Bonnet model will be referred to with subscript $DGB$ for clarity. Their equations of motion are a single divergence of the dual and non-dual Riemann tensor, forming a dually invariant pair of equations of motion $\bar{E}^{\rho\lambda\sigma}_{DRR} = \partial_\lambda {R}^{\omega\rho\lambda\sigma}$ and $\bar{E}^{\rho\lambda\sigma}_{DGB} = \partial_\lambda \mathcal{R}^{\omega\rho\lambda\sigma}$, where the bar represents that the single divergence equation of motion follows from integration by parts of the Euler-Lagrange equation necessary for a gauge invariant, conserved, symmetric and trace-free energy-momentum tensor. These equations of motion correspond to the Maxwell-like higher spin gauge theories for $N = M = 2$, models that have been well explored in the literature \cite{francia2002,francia2012,bekaert2015}. The dually invariant Lagrangian densities remain unchanged $\mathcal{L}_{DRR} =  - \frac{1}{8} ( {R}_{\mu\nu\alpha\beta} {R}^{\mu\nu\alpha\beta} + \mathcal{R}_{\mu\nu\alpha\beta} \mathcal{R}^{\mu\nu\alpha\beta} )$ and $\mathcal{L}_{DGB} = - \frac{1}{4} \mathcal{R}_{\mu\nu\alpha\beta} {R}^{\mu\nu\alpha\beta}$. The energy-momentum tensor $T^{\omega\nu}_{DRR} = R^{\omega\rho\lambda\sigma} R^\nu_{\ \rho\lambda\sigma} -\frac{1}{4}  \eta^{\omega\nu} {R}_{\mu\gamma\alpha\beta} {R}^{\mu\gamma\alpha\beta}$ can also be expressed in dually symmetric form by re-writing the term proportional to Minkowski via the identity in Equation (\ref{drriden}) ($- \frac{1}{4} \eta^{\omega\nu} {R}_{\mu\gamma\alpha\beta} {R}^{\mu\gamma\alpha\beta} 
= - \frac{1}{2} R^{\omega\rho\lambda\sigma} R^\nu_{\ \rho\lambda\sigma} 
- \frac{1}{2} \mathcal{R}^{\omega\rho\lambda\sigma} \mathcal{R}^{\nu}_{\ \rho\lambda\sigma}$), which yields,

\begin{equation}
T^{\omega\nu}_{DRR} = \frac{1}{2} ( R^{\omega\rho\lambda\sigma} R^\nu_{\ \rho\lambda\sigma} - \mathcal{R}^{\omega\rho\lambda\sigma} \mathcal{R}^{\nu}_{\ \rho\lambda\sigma}) ,
\end{equation}

which is analogous to the $MNH$ energy-momentum tensor $T^{\mu\nu}_{MNH}$ in Equation (\ref{TMNH}). Therefore the linearized gravity model can be expressed completely in dual invariant form by considering the equation of motion dual to the linearized Gauss-Bonnet equation of motion, namely the second Bianchi identity; thus the linearized gravity model from $\mathcal{L}_{DGB}$ and $\mathcal{L}_{DRR}$ is analogous to the complete theory of electrodynamics. The equations of motion consisting of a single derivative of the linearized Riemann and dual Riemann tensors is a consequence of the requirement that the models be completely gauge invariant under the spin-2 gauge transformation (linearized diffeomorphisms), as well as have energy-momentum tensors that are conserved, symmetric and trace-free. The $N = M = 1$ and $N = M =2$ dual formulations will be summarized in the following section.

\section{Complete dual models for $N = M = 1$ and $N = M = 2$ \label{s6}}

The two models derived in \cite{baker2019}, namely electrodynamics and the linearized Gauss-Bonnet model have been expressed in dual form. This involved considering the general Lagrangian density $\mathcal{L} = \tilde{a} R_{\mu\nu\alpha\beta} R^{\mu\nu\alpha\beta} + \tilde{b} R_{\mu\nu} R^{\mu\nu} + \tilde{c} R^2$ derived in \cite{baker2019} for $N = M = 2$, of which $R_{\mu\nu\alpha\beta} R^{\mu\nu\alpha\beta}$ ($DRR$) is required to have an equation of motion dual to the second Bianchi identity, which was derived from the Euler-Lagrange equation of the linearized Gauss-Bonnet model ($DGB$). The electrodynamics model derived from $N = M = 1$, consisting of the complete set of Maxwell's equations in the Euler-Lagrange equation, was expressed in terms of the conventional Lagrangian which corresponds to the non-homogenous set of Maxwell's equations ($MNH$), and the Lagrangian which corresponds to the homogenous set of Maxwell's equation ($MH$). A summary of the models for ($MH$), ($MNH$), ($DGB$) and ($DRR$) is presented below,

\begin{center}
\begin{tabular}{ |c|c|c| } 
 \hline
 {\bf{Equation}} & {\bf{Model $N = M = 1$}} & {\bf{Model $N = M = 2$}} \\ 
 \hline
 Non-Homogenous $E^A$ & ${E}^\rho_{MNH} = \partial_\sigma F^{\sigma\rho}$ & $\bar{E}^{\rho\lambda\sigma}_{DRR} = \partial_\lambda {R}^{\omega\rho\lambda\sigma}$ \\ 
 \hline
 Homogenous $E^A$ (I) & $E^\rho_{MH} = \partial_\sigma \mathcal{F}^{\sigma\rho} 
$ & $\bar{E}^{\rho\lambda\sigma}_{DGB} =  \partial_\lambda \mathcal{R}^{\omega\rho\lambda\sigma} $ \\ 
 \hline
 Homogenous $E^A$ (II) & $\partial_\sigma F_{\alpha\beta}+\partial_\beta F_{\sigma\alpha}+\partial_\alpha F_{\beta\sigma} = 0$ & $
\partial_\omega R_{\mu\nu\alpha\beta} + \partial_\mu R_{\nu\omega\alpha\beta} + \partial_\nu R_{\omega\mu\alpha\beta}
= 0$ \\ 
 \hline
Non-Homogenous $\mathcal{L}$ & $\mathcal{L}_{MNH} = - \frac{1}{8} (F_{\mu\nu} F^{\mu\nu} - \mathcal{F}_{\mu\nu} \mathcal{F}^{\mu\nu})$ & $\mathcal{L}_{DRR} = - \frac{1}{8} ( {R}_{\mu\nu\alpha\beta} {R}^{\mu\nu\alpha\beta} + \mathcal{R}_{\mu\nu\alpha\beta} \mathcal{R}^{\mu\nu\alpha\beta} )$ \\ 
 \hline
 Homogenous $\mathcal{L}$ & $\mathcal{L}_{MH} = - \frac{1}{4} \mathcal{F}_{\mu\nu} {F}^{\mu\nu}$ & $\mathcal{L}_{DGB} = - \frac{1}{4} \mathcal{R}_{\mu\nu\alpha\beta} {R}^{\mu\nu\alpha\beta}$ \\ 
 \hline
 Non-Homogenous $T^{\omega\nu}$ & $T^{\omega\nu}_{MNH} = \frac{1}{2} [{F}^{\omega\alpha}  {F}^{\nu}_{\ \alpha}   + \mathcal{F}^{\omega\alpha} \mathcal{F}^{\nu}_{\ \alpha}]$ & $T^{\omega\nu}_{DRR} = \frac{1}{2} [ R^{\omega\rho\lambda\sigma} R^\nu_{\ \rho\lambda\sigma} - \mathcal{R}^{\omega\rho\lambda\sigma} \mathcal{R}^{\nu}_{\ \rho\lambda\sigma}]$ \\ 
 \hline
 Homogenous $T^{\omega\nu}$ & $T^{\omega\nu}_{MH} = \mathcal{F}^{\omega\alpha} F^\nu_{\ \alpha} - \frac{1}{4} \eta^{\omega\nu} \mathcal{F}^{\alpha\beta} F_{\alpha\beta}$ & $T^{\omega\nu}_{DGB} =
\mathcal{R}^{\omega\alpha\beta\lambda} {R}^{\nu}_{\ \alpha\beta\lambda} 
- \frac{1}{4} \eta^{\omega\nu} \mathcal{R}^{\mu\gamma\alpha\beta} {R}_{\mu\gamma\alpha\beta}$ \\ 
 \hline
\end{tabular}
\end{center}

where $E^A$ represents a general Euler-Lagrange equation of motion. Note that the sign change in the Lagrangian and energy-momentum tensor from $N = M = 1$ to $N = M = 2$ is a result of the Minkowski metric causing negative generalized Kronecker deltas, of which $\mathcal{F}_{\mu\nu} \mathcal{F}^{\mu\nu}$ terms have one and $\mathcal{R}_{\mu\nu\alpha\beta} \mathcal{R}^{\mu\nu\alpha\beta}$ terms have two such contributions. The homogenous equations of motion have been expressed both (I) in terms of the dual field strength tensor and (II) in terms of the expanded form (second Bianchi identity). From the summary above, every Lagrangian, equation of motion and energy-momentum tensor is completely and independently invariant under the gauge transformations $A_\mu' = A_\mu + \partial_\mu \phi$ and $h_{\mu\nu}' = h_{\mu\nu} + \partial_\mu \xi_\nu + \partial_\nu \xi_\mu$, thus complete gauge invariance derived in \cite{baker2019} is maintained. Each Lagrangian density and energy-momentum tensor can be expressed in independently dual invariant form. Each model has a pair of dually invariant equations of motion, namely Maxwell's equations $E^\rho_{MH} \leftrightarrow E^\rho_{MNH}$ for $N = M = 1$ and the linearized gravity model $\bar{E}^{\rho\lambda\sigma}_{DRR} \leftrightarrow \bar{E}^{\rho\lambda\sigma}_{DRR}$ for $N = M = 2$. The two homogenous halves of the models ($MH$ and $DBG$) have common form between Lagranians $\mathcal{L}_{MH} \ , \ \mathcal{L}_{DGB}$, equations of motion ${E}^\rho_{MH} \ , \ \bar{E}^{\rho\lambda\sigma}_{DGB}$, and energy-momentum tensors $T^{\omega\nu}_{MH} \ , \ T^{\omega\nu}_{DGB}$. Similarly, the two non-homogenous halves of the models ($MNH$ and $DRR$) have common form between Lagrangians $\mathcal{L}_{MNH} \ , \ \mathcal{L}_{DRR}$, equations of motion ${E}^\rho_{MNH} \ , \ \bar{E}^{\rho\lambda\sigma}_{DRR}$, and energy-momentum tensors $T^{\omega\nu}_{MNH} \ , \ T^{\omega\nu}_{DRR}$. Due to the common form of the two models we can generalize the complete gauge invariant dual formulations to an arbitrary field strength tensor $S$ for $N = M = n$, with equations of motion corresponding to the Maxwell-like higher spin gauge theories \cite{francia2002,francia2012,bekaert2015}, which will be presented in the following section.

\section{Generalization to Maxwell-like higher spin gauge theories \label{s7}}

Gauge invariant curvature (field strength) tensors have long been generalized to all spin-$n$ models, representing the case $N = M = n$. The models for $N = M = 1$ and $N = M = 2$ are built using the spin-1 (Maxwell field strength ${F}^{\omega\alpha}$) and spin-2 (linearized Riemann ${R}^{\mu\nu\alpha\beta}$) curvature tensors. For example, in the spin-3 and spin-4 cases, respectively, we have the field strength (curvature) tensors \cite{damour1987,sorokin2005},

\begin{equation}
\fl
\eqalign{
S^{[\t \n][\k \m] [\x \g]} {}&=  \p^{\x} \p^{\t} \p^{\k} \phi^{\g \m \n}  +   \p^{\t} \p^{\m} \p^{\g} \phi^{\k \x \n} +   \p^{\k} \p^{\n} \p^{\g} \phi^{\x \t \m} +  \p^{\x} \p^{\m} \p^{\n} \phi^{\k \t \g} 
\\
&- \p^{\g} \p^{\m} \p^{\n} \phi^{\x \t \k} - \p^{\x} \p^{\t} \p^{\m} \phi^{\k \n \g} 
 - \p^{\k} \p^{\x} \p^{\n} \phi^{\t \m \g}  - \p^{\t} \p^{\k} \p^{\g} \phi^{\x \m \n}   ,
}
\end{equation}

\begin{equation}
\fl
\eqalign{
S^{[\a \b] [\g \x] [\m \k] [\n \t]} {}&=  \p^{ \a}\p^{\g }\p^{ \m}\p^{ \n} \phi^{ \b \x \k \t}   
+	 \p^{ \a}\p^{\g }\p^{ \k }\p^{ \t}  \phi^{ \b \x \m \n}  	
+	 \p^{ \a}\p^{\m }\p^{ \x }\p^{ \t}  \phi^{ \b \k \g \n}	
+	 \p^{ \a}\p^{\n }\p^{ \x }\p^{ \k}  \phi^{ \b \t \g \m}
\\
  & 
+ 	  \p^{ \g}\p^{\m }\p^{ \b }\p^{ \t}  \phi^{ \x \k \a \n}
 + 	  \p^{ \g}\p^{\n }\p^{ \b }\p^{ \k}  \phi^{ \x \t \a \m} 	
+ 	  \p^{ \m}\p^{\n }\p^{ \b }\p^{ \x}  \phi^{ \k \t \a \g}
    +     \p^{ \b}\p^{\x }\p^{ \k}\p^{ \t}  \phi^{ \a \g \m \n} 
\\
&
-    \p^{ \a}\p^{\g }\p^{ \m}\p^{ \t}   \phi^{ \b \x \k \n}   
-		 \p^{ \a}\p^{\g }\p^{ \n}\p^{ \k}   \phi^{ \b \x \t \m}  
- 	 \p^{ \a}\p^{\m }\p^{ \n}\p^{ \x}   \phi^{ \b \k \t \g}   
 -	 \p^{ \g}\p^{\m }\p^{ \n}\p^{ \b}   \phi^{ \x \k \t \a}   
\\   
 &-    	 \p^{ \a}\p^{\x }\p^{ \k }\p^{ \t}  \phi^{ \b \g \m \n}   
 -  	 \p^{ \g}\p^{\b }\p^{ \k }\p^{ \t}  \phi^{ \x \a \m \n}  
 -	 \p^{ \m}\p^{\b }\p^{ \x }\p^{ \t}  \phi^{ \k \a \g \n} 	 
-	 \p^{ \n}\p^{\b }\p^{ \x }\p^{ \k}  \phi^{ \t \a \g \m}  .
}
\end{equation}

These tensors are generalizations of the linearized Riemann tensors with $n$ pairs of antisymmetric indices that are symmetric under interchange, with totally symmetric $\phi$ for all $n$. They are exactly invariant under the spin-$n$ gauge transformations \cite{sorokin2005}. For spin-3 and spin-4, respectively, these gauge transformations are,

\begin{equation}
\phi_{\a \b \r}' = \phi_{\a \b \r} + \p_{\a} \lambda_{\b \r} + \p_{\b} \lambda_{\a \r} + \p_{\r} \lambda_{\a \b}  ,
\end{equation}

\begin{equation}
\phi_{ \b \x \k \t}'  = \phi_{ \b \x \k \t} +  \p_{\b} \lambda_{ \x \k \t} + \p_{\x }\lambda_{\b \k \t} + \p_{ \k}\lambda_{\b \x  \t} + \p_{\t} \lambda_{\b \x \k } ,
\end{equation}

where the gauge parameters $\lambda$ are totally symmetric for all $n$. The general form of the equations of motion in the previous section are a single divergence of a curvature tensor; models that have been well worked out in the literature for the spin-$n$ curvature tensors, known as the Maxwell-like higher spin gauge theories \cite{francia2002,francia2012,bekaert2015}. The analogy comes from the single divergence of the spin-1 curvature tensor in the case of electromagnetic theory. The dual formulation of these higher spin models has already been explored to some degree \cite{bekaert2003,hinterbichler2016}. In addition, scalars built from the contraction of these higher spin curvature tensors have been considered as exactly gauge invariant Lagrangian densities \cite{francia2010}. Using the curvature tensors of higher spin gauge theories we therefore can build completely dual and gauge invariant models in the analogous form of the $N = M = 1$ and $N = M = 2$ cases summarized in Section \ref{s6}. The general form for the homogenous $H$ and non-homogenous $NH$ equations can be expressed using a general spin-$n$ field strength tensor $S$ and its dual $\mathcal{S}$, suppressing contracted indices, as,

\begin{center}
\begin{tabular}{ |c|c|c| } 
 \hline
 {\bf{Equation}} & {\bf{Model $N = M = n$}}  \\ 
 \hline
 Non-Homogenous $E^A$ & ${E}^A_{NH} = \partial S^{A}$  \\ 
 \hline
 Homogenous $E^A$  & $E^A_{H} = \partial \mathcal{S}^{A} 
$  \\ 
 \hline
Non-Homogenous $\mathcal{L}$ & $\mathcal{L}_{NH} = - \frac{1}{8} (S S \pm \mathcal{S} \mathcal{S})$  \\ 
 \hline
 Homogenous $\mathcal{L}$ & $\mathcal{L}_{H} = - \frac{1}{4} S \mathcal{S}$  \\ 
 \hline
 Non-Homogenous $T^{\omega\nu}$ & $T^{ab}_{NH} = \frac{1}{2} [{S}^{\omega}  {S}^{\nu}   \mp \mathcal{S}^{\omega} \mathcal{S}^{\nu}]$  \\ 
 \hline
 Homogenous $T^{\omega\nu}$ & $T^{ab}_{H} = \mathcal{S}^{\omega} S^\nu - \frac{1}{4} \eta^{\omega\nu} \mathcal{S} S $ \\ 
 \hline
\end{tabular}
\end{center}

where the $\pm$ and $\mp$ refer to odd $n$ models (top sign) and even $n$ models (bottom sign) due to the generalized Kronecker delta in Minkowski spacetime. It is worth noting that the four invariants for $N=M=1$ and $N=M=2$ are associated to (omitting indices) the wedge product between the differential forms representing the field strength tensor of electrodynamics ${\it{F}}$ and its dual $\star {\it{F}}$, and the differential forms representing the Riemann tensor ${\it{R}}$ and its dual $\star {\it{R}}$. Roughly speaking these correspond to the 4 invariants as $\mathcal{L}_{MH} \propto {\it{F}} \wedge {\it{F}}$, $\mathcal{L}_{MNH} \propto {\it{F}} \wedge \star {\it{F}}$, $\mathcal{L}_{DGB} \propto {\it{R}} \wedge {\it{R}}$ and $\mathcal{L}_{DRR} \propto {\it{R}} \wedge \star {\it{R}}$. In general the conjecture can be made that the higher spin models for $N = M = n$ will consider of invariants of field strength $S$, its dual $\mathcal{S}$, and differential form ${\it{S}}$, the presumed general form of the homogenous $H$ and non-homogenous $NH$ invariants will be $\mathcal{L}_{H} \propto {\it{S}} \wedge {\it{S}}$ and $\mathcal{L}_{NH} \propto {\it{S}} \wedge \star {\it{S}}$.

If we are to consider the combined action $\mathcal{L}_{H} + \mathcal{L}_{NH} = - \frac{1}{4} S \mathcal{S} - \frac{1}{8} (S S \pm \mathcal{S} \mathcal{S})$ for a general model, there are two possible generalizations that can be noted. First is the possibility to have a linear combination of the field strength and dual factored $\mathcal{L}_{H} + \mathcal{L}_{NH}  \approx \pm \frac{1}{8} (S + \mathcal{S}) (S + \mathcal{S})$, with signs depending on the particular $S$. The second is that, in the case of electrodynamics where we know the specific fields in each component, this generalization produces $\mathcal{L}_{MH} + \mathcal{L}_{MNH} \approx \frac{1}{2} B^2 - \frac{1}{2} E^2 + \vec{B} \cdot \vec{E}$. These two definitions are strikingly similar to the law of cosines where $c^2 = (\vec{a} -  \vec{b})\cdot(\vec{a} - \vec{b}) = a^2 + b^2 - 2 \vec{a} \cdot \vec{b}$. This similarly, if any meaningful relationship exists, has not been determined.

\section{Conclusions \label{s8}}

In \cite{baker2019}, a procedure was developed for building completely gauge invariant models by imposing gauge invariance and using Noether's first theorem for general Lagrangian densities of $N$ order of derivatives and $M$ rank of tensor potential.  For $N = M = 1$ electrodynamics was uniquely derived, and for $N = M = 2$ linearized Gauss-Bonnet gravity was uniquely derived. Both of these models have the property of complete gauge invariance of the Lagrangian, equation of motion and energy-momentum tensor. The energy momentum tensors are gauge invariant, symmetric, trace-free and conserved. In the recent literature these models were to prove that the Noether and Hilbert energy-momentum tensor are not, in general, equivalent \cite{baker2021a}.

In order to further investigate this relationship, electrodynamics and linearized Gauss-Bonnet gravity were expressed in their respective dual formulations. The Gauss-Bonnet model, conventionally claimed to have simply no equation of motion, in fact has the second Bianchi identity as its equation of motion, analogous to the homogenous half of Maxwell's equations in electrodynamics. In order to introduce the equation of motion dual to this expression, the linearized Riemann-Riemann Lagrangian was introduced, whose dual formulation was analogous to the non-homogenous half of electrodynamics. In this presentation, the electrodynamics and linearized gravity models have internal dual symmetries in their Lagrangians and energy-momentum tensors, and have equations of motion that are dual between the homogenous and non-homogenous halves of the models. The energy-momentum tensors are all gauge invariant, symmetric, trace-free and conserved, with the additional property of dual invariance being explicit in this formulation.  

The dual formulation shared by these two models allows for their homogenous and non-homogenous halves to be expressed in a more general framework. The equations of motion of this general framework correspond to the Maxwell-like higher spin gauge theories built from the spin-$n$ curvature tensors. These models are completely gauge invariant in the same manner as the electrodynamics and linearized Gauss-Bonnet gravity cases. In addition they are dually invariant analogous to the results in this article. Obtaining physical models which have some uniqueness criteria that separate themselves from other possible equations has been one of the focuses of theoretical physics in recent decades. Electrodynamics, perhaps the most successful model in physics, has a plethora of such properties: complete gauge invariance, conformal invariance, dual invariance, a trace-free and symmetric energy-momentum tensor, just to name a few. What we have shown is that these uniqueness properties can be generalized to the higher spin (Maxwell-like) gauge theories, where the linearized Gauss-Bonnet gravity model is the $N = M = 2$ analogue to the homogenous half of Maxwell's equations. Recent research has brought great renewed interest in the Gauss-Bonnet gravity model \cite{glavan2020}, as it has been claimed to provide `new' dynamical predictions that explain astronomical observations to a higher degree of accuracy. Due to this, the $N = M = 2$ model can be applied to some of these observations to see if it too can better explain some observed phenomena; this application is the subject of future work.

\section{Acknowledgement}

We are grateful to N. Kiriushcheva and S. Kuzmin for numerous discussions and suggestions during the preparation of this paper.

\section{Bibliography}

\bibliographystyle{unsrt}
\bibliography{IJMPBib_B_2020}

\begin{thebibliography}{10}

\bibitem{baker2019}
M.R. Baker and S.~Kuzmin.
\newblock A connection between linearized gauss--bonnet gravity and classical
  electrodynamics.
\newblock {\em International Journal of Modern Physics D}, 28(07):1950092,
  2019.

\bibitem{noether1918}
E.~Noether.
\newblock Invariante variationsprobleme.
\newblock {\em König. Gesellsch. d. Wiss. zu Göttingen, Math.-Phys. Klasse},
  pages 235--257, 1918.

\bibitem{kosmann2011noether}
Y.~Kosmann-Schwarzbach.
\newblock {\em The Noether Theorems: Invariance and Conservation Laws in the
  Twentieth Century - translation of E. Noether's `Invariante
  Variationsprobleme'}.
\newblock Sources and Studies in the History of Mathematics and Physical
  Sciences. Springer New York, 2010.

\bibitem{baker2021a}
M.R. Baker, N.~Kiriushcheva, and S.~Kuzmin.
\newblock {Noether and Hilbert} (metric) energy-momentum tensors are not, in
  general, equivalent.
\newblock {\em Nuclear Physics B}, 962:115240, 2021.

\bibitem{francia2002}
D.~Francia and A.~Sagnotti.
\newblock Free geometric equations for higher spins.
\newblock {\em Physics Letters B}, 543(3-4):303--310, 2002.

\bibitem{francia2012}
D.~Francia.
\newblock Generalized connections and higher spin equations.
\newblock {\em Classical and Quantum Gravity}, 29(24):245003, 2012.

\bibitem{bekaert2015}
X.~Bekaert, N.~Boulanger, and D.~Francia.
\newblock Mixed-symmetry multiplets and higher-spin curvatures.
\newblock {\em Journal of Physics A: Mathematical and Theoretical},
  48(22):225401, 2015.

\bibitem{eguchi1980}
T.~Eguchi, P.B. Gilkey, and A.J. Hanson.
\newblock Gravitation, gauge theories and differential geometry.
\newblock {\em Physics reports}, 66(6):213--393, 1980.

\bibitem{gauss1827}
C.F. Gauss.
\newblock {\em Disquisitiones generales circa superficies curvas}, volume~1.
\newblock Typis Dieterichianis, 1827.

\bibitem{bonnet1848}
O.~Bonnet.
\newblock {\em M{\'e}moire sur la th{\'e}orie g{\'e}n{\'e}ral des surfaces}.
\newblock Bachelier, 1848.

\bibitem{dyck1890}
W.~Dyck.
\newblock Beitr{\"a}ge zur analysis situs.
\newblock {\em Mathematische Annalen}, 37(2):273--316, 1890.

\bibitem{hopf1926}
H.~Hopf.
\newblock Vektorfelder inn-dimensionalen mannigfaltigkeiten.
\newblock {\em Mathematische Annalen}, 96(1):225--249, 1926.

\bibitem{chern1944}
S.~Chern.
\newblock A simple intrinsic proof of the gauss-bonnet formula for closed
  riemannian manifolds.
\newblock {\em Annals of mathematics}, pages 747--752, 1944.

\bibitem{allendoerfer1940}
C.B. Allendoerfer.
\newblock The euler number of a riemann manifold.
\newblock {\em American Journal of Mathematics}, 62(1):243--248, 1940.

\bibitem{lanczos1938}
C.~Lanczos.
\newblock A remarkable property of the riemann-christoffel tensor in four
  dimensions.
\newblock {\em Annals of Mathematics}, pages 842--850, 1938.

\bibitem{wu2008}
H.-H. Wu.
\newblock Historical development of the gauss-bonnet theorem.
\newblock {\em Science in China Series A: Mathematics}, 51(4):777--784, 2008.

\bibitem{ray1978}
J.R. Ray.
\newblock A variational derivation of the bach--lanczos identity.
\newblock {\em Journal of Mathematical Physics}, 19(1):100--102, 1978.

\bibitem{zwiebach1985}
B.~Zwiebach.
\newblock Curvature squared terms and string theories.
\newblock {\em Physics Letters B}, 156(5-6):315--317, 1985.

\bibitem{boulware1985}
D.G. Boulware and S.~Deser.
\newblock String-generated gravity models.
\newblock {\em Physical Review Letters}, 55(24):2656, 1985.

\bibitem{myers1987}
R.C. Myers.
\newblock Higher-derivative gravity, surface terms, and string theory.
\newblock {\em Physical Review D}, 36(2):392, 1987.

\bibitem{charmousis2002}
C.~Charmousis and J.F. Dufaux.
\newblock General gauss--bonnet brane cosmology.
\newblock {\em Classical and Quantum Gravity}, 19(18):4671, 2002.

\bibitem{cherubini2002}
C.~Cherubini, D.~Bini, S.~Capozziello, and R.~Ruffini.
\newblock Second order scalar invariants of the riemann tensor: applications to
  black hole spacetimes.
\newblock {\em International Journal of Modern Physics D}, 11(06):827--841,
  2002.

\bibitem{granda2012}
L.N. Granda.
\newblock Late time cosmological scenarios from scalar field with gauss bonnet
  and non-minimal kinetic couplings.
\newblock {\em International Journal of Theoretical Physics}, 51(9):2813--2829,
  2012.

\bibitem{marugame2016}
T.~Marugame.
\newblock Renormalized chern-gauss-bonnet formula for complete
  k{\"a}hler-einstein metrics.
\newblock {\em American Journal of Mathematics}, 138(4):1067--1094, 2016.

\bibitem{benetti2018}
M.~Benetti, S.~Santos~da Costa, S.~Capozziello, J.S. Alcaniz, and
  M.~De~Laurentis.
\newblock Observational constraints on gauss--bonnet cosmology.
\newblock {\em International Journal of Modern Physics D}, 27(08):1850084,
  2018.

\bibitem{glavan2020}
D.~Glavan and C.~Lin.
\newblock Einstein-gauss-bonnet gravity in four-dimensional spacetime.
\newblock {\em Physical Review Letters}, 124(8):081301, 2020.

\bibitem{fernandes2020}
P.G.S. Fernandes.
\newblock Charged black holes in ads spaces in 4d einstein gauss-bonnet
  gravity.
\newblock {\em Physics Letters B}, page 135468, 2020.

\bibitem{heaviside1894}
O.~Heaviside.
\newblock {\em Electrical papers}, volume~2.
\newblock 1894.

\bibitem{minkowski1909}
H.~Minkowski.
\newblock Raum und zeit.
\newblock {\em Physikalische Zeitschrift}, 10:104--111, 1909.

\bibitem{einstein1916}
A.~Einstein.
\newblock A new formal interpretation of maxwell's field equations of
  electrodynamics.
\newblock {\em Sitzungsber. Preuss. Akad. Wiss. Berlin, Math.-Phys Klasse},
  pages 184--188, 1916.

\bibitem{padmanabhan2008}
T.~Padmanabhan.
\newblock From gravitons to gravity: Myths and reality.
\newblock {\em International Journal of Modern Physics D}, 17:367--398, 2008.

\bibitem{magnano2002}
G.~Magnano and L.M. Sokolowski.
\newblock Symmetry properties under arbitrary field redefinitions of the metric
  energy--momentum tensor in classical field theories and gravity.
\newblock {\em Classical and Quantum Gravity}, 19(2):223, 2002.

\bibitem{dewit1980}
B.~de~Wit and D.Z. Freedman.
\newblock Systematics of higher-spin gauge fields.
\newblock {\em Physical Review D}, 21(2):358, 1980.

\bibitem{cameron2012}
R.P. Cameron and S.M. Barnett.
\newblock Electric--magnetic symmetry and noether's theorem.
\newblock {\em New Journal of Physics}, 14(12):123019, 2012.

\bibitem{bliokh2013}
K.Y. Bliokh, A.Y. Bekshaev, and F.~Nori.
\newblock Dual electromagnetism: helicity, spin, momentum and angular momentum.
\newblock {\em New Journal of Physics}, 15(3):033026, 2013.

\bibitem{gibbons1995}
G.W. Gibbons and D.A. Rasheed.
\newblock Electric-magnetic duality rotations in non-linear electrodynamics.
\newblock {\em Nuclear Physics B}, 454(1-2):185--206, 1995.

\bibitem{gaillard1998}
M.K. Gaillard and B.~Zumino.
\newblock Self-duality in nonlinear electromagnetism.
\newblock In {\em Supersymmetry and quantum field theory}, pages 121--129.
  Springer, 1998.

\bibitem{kuzenko2013}
S.M. Kuzenko.
\newblock Duality rotations in supersymmetric nonlinear electrodynamics
  revisited.
\newblock {\em Journal of High Energy Physics}, 2013(3):153, 2013.

\bibitem{maeda2008}
H.~Maeda and M.~Nozawa.
\newblock Generalized misner-sharp quasilocal mass in einstein-gauss-bonnet
  gravity.
\newblock {\em Physical Review D}, 77(6):064031, 2008.

\bibitem{kling2005}
T.P. Kling and B.~Keith.
\newblock The bianchi identity and weak gravitational lensing.
\newblock {\em Classical and Quantum Gravity}, 22(14):2921, 2005.

\bibitem{besselhagen1921}
E.~Bessel-Hagen.
\newblock {\"U}ber die erhaltungss{\"a}tze der elektrodynamik.
\newblock {\em Mathematische Annalen}, 84(3-4):258--276, 1921.

\bibitem{jackiw1978}
R.~Jackiw.
\newblock Gauge-covariant conformal transformations.
\newblock {\em Physical Review Letters}, 41(24):1635, 1978.

\bibitem{deser2010}
S.~Deser.
\newblock Gravity from self-interaction redux.
\newblock {\em General Relativity and Gravitation}, 42(3):641--646, 2010.

\bibitem{damour1987}
T.~Damour and S.~Deser.
\newblock “geometry” of spin 3 gauge theories.
\newblock In {\em Annales de l'IHP Physique th{\'e}orique}, volume~47, pages
  277--307, 1987.

\bibitem{sorokin2005}
D.~Sorokin.
\newblock Introduction to the classical theory of higher spins.
\newblock In {\em AIP Conference Proceedings}, volume 767, pages 172--202.
  American Institute of Physics, 2005.

\bibitem{bekaert2003}
X.~Bekaert and N.~Boulanger.
\newblock On geometric equations and duality for free higher spins.
\newblock {\em Physics Letters B}, 561(1-2):183--190, 2003.

\bibitem{hinterbichler2016}
K.~Hinterbichler and A.~Joyce.
\newblock Manifest duality for partially massless higher spins.
\newblock {\em Journal of High Energy Physics}, 2016(9):141, 2016.

\bibitem{francia2010}
D.~Francia.
\newblock String theory triplets and higher-spin curvatures.
\newblock {\em Physics Letters B}, 690(1):90--95, 2010.

\end{thebibliography}

\newpage

\end{document}